\documentclass[english,preprint,12pt]{aastex}

\slugcomment{}
\shorttitle{}
\shortauthors{}

\begin{document}

\title{Viscous diffusion and photoevaporation of stellar disks}

\author{Isamu Matsuyama}

\affil{Department of Astronomy and Astrophysics,University of Toronto,
Toronto,
ON M5S 3H8, Canada}

\email{isamu@astro.utoronto.ca}

\author{Doug Johnstone}

\affil{National Research Council Canada, Herzberg Institute of
Astrophysics,
5071 West Saanich Road, Victoria, BC V9E 2E7, Canada}

\email{doug.johnstone@nrc.ca}

\and{}

\author{Lee Hartmann}

\affil{Harvard-Smithsonian Center for Astrophysics, 60 Garden Street,
Cambridge,
MA 02138}

\email{lhartmann@cfa.harvard.edu}

\begin{abstract}
The evolution of a stellar disk under the influence of viscous
evolution,
photoevaporation from the central source, and photoevaporation by
external stars is studied. We take the typical parameters of TTSs
and the Trapezium Cluster conditions. The photoionizing flux from the
central source is assumed to arise both from the quiescent star and 
accretion shocks at the base of stellar magnetospheric columns, along
which 
material from the disk accretes. The accretion flux is calculated 
self-consistently from the accretion mass loss rate. We find that
the disk cannot be entirely removed using only viscous evolution
and photoionization from the disk-star accretion shock. However,
when FUV photoevaporation by external massive stars is included the
disk is removed in \( 10^{6}-10^{7} \)yr; and
when EUV photoevaporation by external massive
stars is included the disk is removed in \( 10^{5}-10^{6} \)yr.

An intriguing feature of photoevaporation by the central star is the
formation of a gap in the disk at late stages of the disk evolution.
As the gap starts forming, viscous spreading and photoevaporation
work in resonance. When viscous accretion and photoevaporation by
the central star and external massive stars are considered, the disk
shrinks and is truncated at the gravitational radius, where it is
quickly removed by the combination of viscous accretion, viscous
spreading,
photoevaporation from the central source, and photoevaporation by
the external stars. There is no gap formation for disks nearby external
massive stars because the outer annuli are quickly removed by the
dominant EUV flux. On the other hand, at larger, more typical distances
(\( d\gg 0.03 \)pc) from the external stars the flux is FUV dominated.
As a consequence, the disk is efficiently evaporated at two different
locations; forming a gap during the last stages of the disk evolution. 
\end{abstract}


\section{Introduction}

The Hubble Space Telescope (HST) has provided clear evidence of gas
disks surrounding young stars in the Orion Nebula. Narrow band images
reveal circumstellar disks seen in silhouette against either the
background
nebular light or the proplyd's own ionization front \citep{orion
observations}.
These disks have been identified as {}``evaporating'' by \citet*{j98}.
Theoretically, disks should be ubiquitous. Any breaking of the spherical
symmetry of the protostellar collapse will result in in-falling material
being deflected from the radial direction, and disks forming around
the central stars. Spherical symmetry may be broken either when the
central star core is magnetized, or when the protostellar cloud has
initial angular momentum. Magnetic fields tend to produce large
pseudo-disks;
since the material is not solely rotationally supported
\citep{magnetized}.
Alternatively, even small initial rotational velocities in the
protostellar
cloud produce rotationally supported disks containing most of the
angular momentum of the system \citep{rotation}. For most theoretical
models of the collapse of rotating clouds, the majority of the cloud
material falls first onto the disk. Thus, as the molecular core
collapses
the disk mass increases. However, it is unlikely that the disk mass,
\( M_{d} \), becomes larger than the superior limit, \( M_{max}\sim
0.3M_{\star } \),
where \( M_{\star } \) is the mass of the central star. At this superior
limit the disk becomes gravitationally unstable, angular momentum
is transported outward by spiral density waves, and the disk accretes
material toward the central star at almost the same rate as it is
receiving material from the molecular core \citep{mmax}.

Planet formation is an exciting possible outcome of proto-stellar
disk evolution. The coplanarity and circularity of the planetary orbits
in our Solar System support this notion. Explaining the origin of
the Solar System and extra-solar systems requires an understanding
not only how the disks form; furthermore, we need to understand the
disk evolution. In particular, the disk removal timescale and the
timescale to assemble planets determine the possibility of planet
formation. Shu, Johnstone, \& Hollenbach (1993) proposed
photoevaporation
of the Solar Nebula as the gas removal mechanism that explains the
differences in envelope masses between the gas-rich giants, Jupiter
and Saturn; and the gas-poor giants, Uranus and Neptune. \citet{removal
mechanisms}
generalized the discussion, describing the variety of possible disk
removal mechanisms. The dominant disk removal mechanism at the inner
parts of the disk is viscous accretion onto the central star. However,
this process is incapable of removing the entire disk in a finite
time because the accretion rate decreases as the viscous disk spreads,
and the disk lifetime becomes infinite. Other possible disk removal
mechanisms are planet formation, stellar encounters, stellar winds
or disk winds, and photoevaporation by ultraviolet photons.
\citet{removal mechanisms}
concluded that planet formation is a minor disk removal mechanism,
and that the dominant mechanisms for a wide range of disk sizes are
viscous accretion and photoevaporation, operating in concert within
the disk.

Recently, \citet*{uvswitch} have studied the observational consequences
of the evolution of disks through a combination of photoevaporation
and viscous disk evolution. Their study focused on photoevaporation
due to ultraviolet photons produced in the disk-star accretion shock
under the assumption that the accretion luminosity was constant during
accretion and switched off when the inner disk was cleared. Using
this model \citet{uvswitch} were able to reproduce the observed
millimeter
fluxes of stars with disks as a function of the observed accretion
rate. In this complimentary study, we focus on the physical properties
of the disk under a variety of photoevaporation and viscous scenarios
in order to understand the internal disk evolution. We use a time
dependent \( \alpha  \)-disk model \citep{alpha} with the parameters
of \citet{h98} that are consistent with observed mass accretion rates
in T Tauri stars (TTSs). Photoevaporation by external stars is studied
using the model and parameterization of \citet{j98}, in their study
of the Orion Nebula. Photoevaporation by the central star is modelled
with solutions originally found for high mass stars \citep{pdr} and
normalized to TTSs \citep{evaporation shu}; however, in order to study
disk evolution, approximations for the time dependence of evaporation
are included in the model by estimating the continual change in the
accretion shock emission of ultraviolet photons as the accretion rate
subsides. In agreement with \citet{removal mechanisms} and
\citet{uvswitch},
we show that it is possible to remove the entire disk in a finite
time. However, we show that the rapid removal of the inner disk,
described
by \citet{uvswitch} is not self-consistent. We further show that gaps
in the disk are a natural outcome of the combination of viscous
accretion
and photoevaporation by the central star.

\section{Disk Model}

\subsection{The thin viscous disk}

With viscosity present in the disk, the energy of the shear motions
between annuli is dissipated as heat, and angular momentum is
transported
from annuli with smaller specific angular momenta to annuli with larger
specific angular momenta. If the total angular momentum in the disk
is conserved a minimum energy state is approached as the inner material
moves closer to the central star and the outer material spreads outward,
with the outward transport of specific angular momentum \citep{acc
disks}.
Considering angular momentum and mass conservation of an annulus of
material at a radius, \( R \), in a geometrically thin disk with
viscosity, \( \nu  \), the surface density evolution is described
by

\begin{equation}
\label{sigma evolution}
\frac{\partial \Sigma }{\partial t}=\frac{3}{R}\frac{\partial }{\partial
R}\left[ R^{1/2}\frac{\partial }{\partial R}\left( \nu \Sigma
R^{1/2}\right) \right] ,
\end{equation}
 where \( \Sigma  \) is the surface density and \( t \) is the time.
In this formulation it is assumed that most of the mass is in the
central star; therefore, the self gravity of the disk is ignored.
The explicit evolution of the disk depends on a detailed description
of the viscosity, which is weakly constrained by present observations.
However, a standard approach is to assume that the viscosity is a
function of the sound speed in the disk and the disk thickens. For this
case, we can isolate all our uncertainty about the viscous mechanism 
in a dimensionless
parameter, \( \alpha \leq 1 \), with the standard \( \alpha
\)-prescription
of Shakura \& Sunyaev (1973). We write \begin{equation}
\label{alpha viscosity}
\nu =\alpha c_{s}H,
\end{equation}
 where \( c_{s} \) is the sound speed at the disk mid plane and \( H \)
is the thickness of the disk. With this prescription the viscosity
takes the form\begin{equation}
\label{viscosity}
\nu (R)=\alpha \frac{k}{m_{p}}\left( \frac{1}{GM_{\star }}\right)
^{1/2}R^{3/2}T_{d}(R),
\end{equation}
 where \( m_{p}\sim 2.3m_{H} \) is the mean particle mass, \( M_{\star }
\)
is the central star mass, and \( T_{d} \) is the disk temperature
at the mid plane.

The potential energy of accreting gas is converted to thermal energy
and released as radiation by viscous processes. One half of this
accretion
luminosity is released at the disk surface and the other half is
released
at the accretion shock \citep[see][]{acc disks}. Assuming thermodynamic
equilibrium, the local energy balance at each radius is described
by \begin{equation}
\label{thermal equilibrium}
\frac{3}{8\pi }\frac{GM_{\star }\dot{M}}{R^{3}}\left[ 1-\left(
\frac{R_{\star }}{R}\right) ^{1/2}\right] =\sigma T_{s}^{4},
\end{equation}
where \( \sigma  \) is Stefan-Boltzmann constant and \( T_{s} \)
is the disk surface temperature. For simplicity we ignore the vertical
temperature
gradient and assume $T_d = T_s$. Equation \ref{thermal equilibrium}
implies a temperature distribution
$T_s \propto R^{-3/4}$ for a constant accretion rate
at each radius.  However, the temperature distribution
in the outer disk required to explain observations
is much shallower, approximately $T_s \propto R^{-1/2}$
(Kenyon \& Hartmann 1987).  The reason for this behavior
is that the optically-thick disk photosphere is curved
away from the disk midplane, or flared, allowing radiation
from the central star to dominate the outer disk heating
(Kenyon \& Hartmann 1987).  Flared disk models have been
shown to account for the observed disk emission satisfactorily
(Chiang \& Goldreich 1997; D'Alessio et al. 1998, 1999), and the
dominance of stellar irradiation heating over viscous dissipation
is demonstrated by the detection of silicate emission features,
which arise from the temperature inversion due to external
heating (Calvet et al. 1992; Chiang \& Goldreich 1997).

For the observed disk temperature distribution, \( T_{d}\propto R^{-1/2}
\),
the viscosity takes the simple form, \( \nu (R)\propto R \), and
the surface density evolution (equation \ref{sigma evolution}) has
analytic similarity solutions \citep{viscous disks}. When only viscous
accretion is considered (i.e. in the absence of photoevaporation),
numerical solutions can be compared to the similarity solutions.
Similarity
solutions present an excellent starting condition for studying the
interaction of disk evaporation and viscous evolution for various
reasons. First, in the absence of photoevaporation any initial surface
density distribution will diffuse to the similarity solution in the
viscous diffusion timescale, \( t_{s}=R^{2}/3\nu  \) \citep[see][]{acc
disks},
which is much shorter than the disk removal timescale. Second, we
are interested in the disk evolution well after the initial disk
formation
during which time viscous diffusion should have produced a disk profile
reasonably close to the similarity solution. Third, the formation
of structure in the disk surface density due to photoevaporation occurs
at late stages of the disk evolution, at timescales much longer than
the viscous diffusion timescale.

We assume the observed typical parameters of TTSs. The central star
has effective temperature \( \sim 4000\textrm{K} \), luminosity \( \sim
1L_{\sun } \),
mass \( \sim 0.5M_{\sun } \) \citep{fuv observations}. We adopt the
observed disk temperature distribution \citep{vertical structure},
\( T_{d}=\left( 10\textrm{K}\right) (R/100\textrm{AU})^{-1/2} \).
These parameters are consistent with accretion rate observations of
TTSs \citep{h98}. The initial surface density distribution is
\begin{equation}
\label{initial solution}
\Sigma =\frac{M_{d}(0)}{2\pi R_{0}^{2}}\frac{1}{\left( R/R_{0}\right)
\tau ^{3/2}}e^{-\left( R/R_{0}\right) \tau }
\end{equation}
 where \( M_{d}(0) \) is the initial disk mass, \( R_{0}=10\textrm{AU}
\)
is a radial distance scale, and \( \tau \equiv t/t_{s}(R_{0}) \)
is a dimensionless time variable normalized to the viscous diffusion
timescale: \( t_{s}(R_{0})=R_{0}^{2}/3\nu (R_{0}) \). The physical
meaning of \( R_{0} \) is that a fraction \( 1-e^{-q} \) of the
initial disk mass resides inside \( qR_{0} \). For example, \( \sim 60\%
\)
of the mass resides inside \( R_{0} \) and \( \sim 90\% \) of the
mass resides inside \( 2\, R_{0} \). \citet{h98} estimate \( \alpha \sim
10^{-2} \)
from the observed disk sizes, and they show that the observed variation
in mass accretion rates can be accounted for by initial disk masses
between \( 0.01 \) and \( 0.2M_{\sun } \).

We solve equation \ref{sigma evolution} numerically using a backward
time finite differencing scheme \citep{numerical recipes in c}. The
advantage of this scheme is that it is stable for any combination
of time step \( \delta t \) and radial distance step \( \delta R \),
provided the boundary conditions are well determined. It is necessary
to solve a set of simultaneous linear equations at each time step
given an initial density distribution and two boundary conditions.
The inner boundary is \( R_{min}=10^{-2} \)AU and the outer boundary,
\( R_{max} \), is chosen such that the outer disk edge never reaches
\( R_{max} \). The boundary conditions are chosen such that the total
mass and the total angular momentum in the disk are conserved. In
other words, there is no input of mass or external torques at the
inner or external boundaries. We define the disk edge, \( R_{d} \),
such that the mass between \( R_{d} \) and \( R_{max} \) is less
than a fraction \( 10^{-6} \) of the disk mass. This assures that
the contribution to the disk mass from all the annuli with \( R>R_{d} \)
is negligible.

\subsection{The star disk accretion shock}

The hot continuum emission or ``blue veiling'' present in TTSs,
originally explained as being due to boundary layer emission
(Lynden-Bell \& Pringle 1974; Bertout, Basri, \& Bouvier 1988),
is now thought to arise from accretion shocks at the base
of stellar magnetospheric columns, along which material from
the disk accretes (K\"onigl 1991; Hartmann, Hewett, \& Calvet 1994;
see Hartmann 1998 and references therein).  Here we assume that
half of the accretion luminosity is radiated as hot continuum to write
\begin{equation}
\label{as luminosity}
L_{as}=\frac{GM_{\star }\dot{M}_{d}}{2R_{\star }},
\end{equation}
where $L_{as}$ is the accretion shock luminosity. Magnetospheric
accretion models generally assume $\sim$80\% of
the accretion energy comes out in veiling continuum, but this
difference is small in comparison with the uncertainty in the
characteristic temperature.

The veiling continua of accreting T Tauri stars is not that of
a simple blackbody.  Nevertheless, it appears that the FUV continuum
can be roughly approximated as having a characteristic temperature
$\sim 10,000$K (Johns-Krull, Valenti, \& Linsky 2000;
Gullbring et al. 2000).  There is no constraint on the
EUV fluxes due to interstellar absorption.
In the following we assume that the FUV and EUV fluxes can be
characterized as blackbody emission at $T = 1.5 \times 10^4$K,
which may somewhat overestimate the amount of short-wavelength
radiation.

\subsection{The hot ionized disk atmosphere}

The EUV (\( h\nu >13.6 \)eV) photons from the central star, the
accretion
shock, or the external stars are capable of ionizing hydrogen and
evaporating material from the disk surface. This mechanism affects
the disk surface layer and forms an ionized atmosphere above the thin
viscous disk. \citet{pdr} describe analytical solutions for this
atmosphere
assuming a typical \ion{H}{2} region temperature, \( T_{II} \), of
\( 10^{4} \)K. The equilibrium temperature arises from balance between
heating, due primarily to incident ionizing photons; and cooling,
due primarily to forbidden line radiation. Using Euler's equation
and assuming hydrostatic equilibrium in the z-direction, we can write
the number density of electrons for an isothermal atmosphere as

\begin{equation}
\label{number density 1}
n(R,z)=n_{0}(R,z=0)e^{-z^{2}/2H^{2}},
\end{equation}
 where \( n_{0}(R) \) is the number density at the disk base \( z=0 \)
and \begin{equation}
\label{scale height}
H=c_{sII}\left( \frac{R}{GM_{\star }}\right) ^{1/2}R,
\end{equation}
 is the scale height. In the last equation \begin{equation}
\label{cs II}
c_{sII}=\left( \frac{kT_{II}}{m_{II}}\right) ^{1/2}
\end{equation}
 is the isothermal sound speed of the gas at temperature, \( T_{II} \),
and mean particle mass, \( m_{II}=1.13\times 10^{-24} \) g. The scale
height grows with increasing \( R \) until it becomes equal to \( R \)
at the gravitational radius, \begin{equation}
\label{rg II}
R_{gII}\equiv \frac{GM_{\star }}{c_{sII}^{2}}.
\end{equation}
 The relevance of this radius is not only geometrical, but also
dynamical.
The sum of the kinetic energy and the thermal energy per unit mass
at the gravitational radius is\begin{equation}
\label{dynamical rg}
\frac{1}{2}\Omega
^{2}R_{gII}^{2}+\frac{3kT_{II}}{2m_{II}}=\frac{2GM_{\star }}{R_{gII}},
\end{equation}
 where \( \Omega =(GM_{\star }/R_{gII}^{3})^{1/2} \) is the Keplerian
angular velocity. Gas material has twice the negative energy of the
gravitational potential energy, that is, it has more than enough energy
to escape to infinity as a disk wind at \( R_{gII} \).
\citet{evaporation shu}
show that the gravitational radius for the Solar nebula is at the
orbital distance of Saturn, giving a possible explanation for the
sharp differences in envelope masses between the gas-rich giants,
Jupiter and Saturn, and the gas-poor giants, Uranus and Neptune. Since
the gas material at \( R_{gII} \) already has more than the minimum
energy to escape, some mass loss occurs even inside \( R_{g} \),
and gas particles become gravitationally bound to the central star
at a location between the central star radius, \( R_{\star } \),
and the gravitational radius. We characterize this by assuming that
the effective gravitational radius is \( \beta R_{gII} \), where
\( R_{\star }/R_{gII}<\beta <1 \). An additional factor that makes
\( \beta  \) smaller is the fact that material can also be removed
inside \( R_{gII} \) due to the pressure gradients in the flow. We
use an effective gravitational radius characterized by \( \beta =0.5 \),
i.e. we assume that the gas material in the ionized atmosphere is
gravitationally bound to the central star for \( R<0.5R_{gII} \).

\subsection{The warm neutral disk atmosphere}

The FUV (\( 6<h\nu <13.6\textrm{eV} \)) photons from the central
star, the accretion shock, or external stars; capable of dissociating
\( \textrm{H}_{2} \) and CO; also affect the disk structure. The
FUV photons penetrate the ionized region and create a neutral hydrogen
layer with temperature, \( T_{I}\sim 10^{3}\textrm{K} \). Following
the same arguments for the ionized atmosphere, we can define a
gravitational
radius for the neutral layer as\begin{equation}
\label{rg I}
R_{gI}\equiv \frac{GM_{\star }}{c_{sI}^{2}}.
\end{equation}
 The isothermal sound speed at the neutral layer is given by
\begin{equation}
\label{cs I}
c_{sI}=\left( \frac{kT_{I}}{m_{I}}\right) ^{1/2},
\end{equation}
 where \( m_{I}=1.35m_{H} \) is the mean particle mass per hydrogen
atom.

\section{Disk evolution and photoevaporation}

\subsection{EUV photoevaporation from the central source}

\citet{pdr} found analytic solutions for the photoevaporation mass
loss rate by EUV photons from the central star. These photons are
attenuated by recombined hydrogen atoms and scattering from dust in
the ionized atmosphere; providing a source of diffuse EUV photons.
The ionized atmosphere absorbs a significant fraction of the direct
incident flux, and the diffuse field dominates the flux onto the disk.
Disk material is gravitationally bound inside the gravitational radius,
\( R_{gII} \), and it flows out the disk base at the sound speed,
\( c_{sII} \), outside the gravitational radius. Given the number
density of ionized hydrogen at the disk base, \( n_{0}(R) \), we
can calculate the evaporation rate:

\begin{equation}\label{central dsigmadt}\dot{\Sigma }(R)=\cases{
2m_{II}n_{0}(R)a_{II}, &if $R>\beta R_{gII}$;\cr 0, &otherwise; \cr
}\end{equation}
where the factor of two accounts for photoevaporation from both sides
of the disk. A self-regulating mechanism is established at the disk
base and it is possible to find the number density in this region.
If the number density at the disk base were lower than the equilibrium
value, the diffuse EUV photons would penetrate deeper into the disk,
producing more ionizations, and the number of ionized hydrogen would
increase. On the other hand, if the number density were higher than
the equilibrium value, the recombinations and the scattering from
dust in the ionized atmosphere would prevent the diffuse EUV photons
from reaching the disk base, and the number of ionizations would
decrease.

Assuming ionization equilibrium \citet{pdr} found the number density
at the disk base for the {}``weak'' and {}``strong'' stellar wind
cases. In the weak stellar wind case, the stellar wind ram pressure
is smaller than the thermal pressure for \( R<\beta R_{gII} \) and
the atmosphere is static. Gas material is evaporated at the rate given
by equation \ref{central dsigmadt} outside the gravitational radius.
The dominant flux of EUV photons producing the flow is from the diffuse
field that shines vertically downward onto the disk at the gravitational
radius; therefore, most of the gas is evaporated from this region.
In the strong stellar wind case, the stellar wind ram pressure is
higher than the thermal pressure even outside \( \beta R_{gII} \).
Although disk material evaporates in the vicinity of \( \beta R_{gII}
\),
the dominant flow occurs where the stellar wind ram pressure equals
the thermal pressure, at a characteristic radius, \( R_{w}>\beta R_{gII}
\).
Due to large uncertainties in both the wind mass loss rate, the effects
of collimation in the wind, and the ionization flux from the central
star, it is not clear whether or not the strong wind condition is
met for low mass stars \citep{removal mechanisms}. Here, we only
consider
the weak stellar wind case. The number density at the disk base for
the weak stellar wind case is

\begin{equation}\label{number density} n_{0}(R)=\cases{ 5.7\times
10^{5}\phi_{40}^{1/2}R_{13}^{-3/2}\textrm{ cm}^{-3}, &if $R\leq \beta
R_{gII}$; \cr n_{0}(\beta R_{gII})\left(R/\beta R_{gII}\right) ^{-5/2 },
&otherwise; \cr } \end{equation}
where \( \phi _{40} \) is the ionizing photon luminosity in units
of \( 10^{40}\textrm{s}^{-1} \) and \( R_{13} \) is the radius in
units of \( 10^{13}\textrm{cm} \).

\begin{figure}

\plottwo{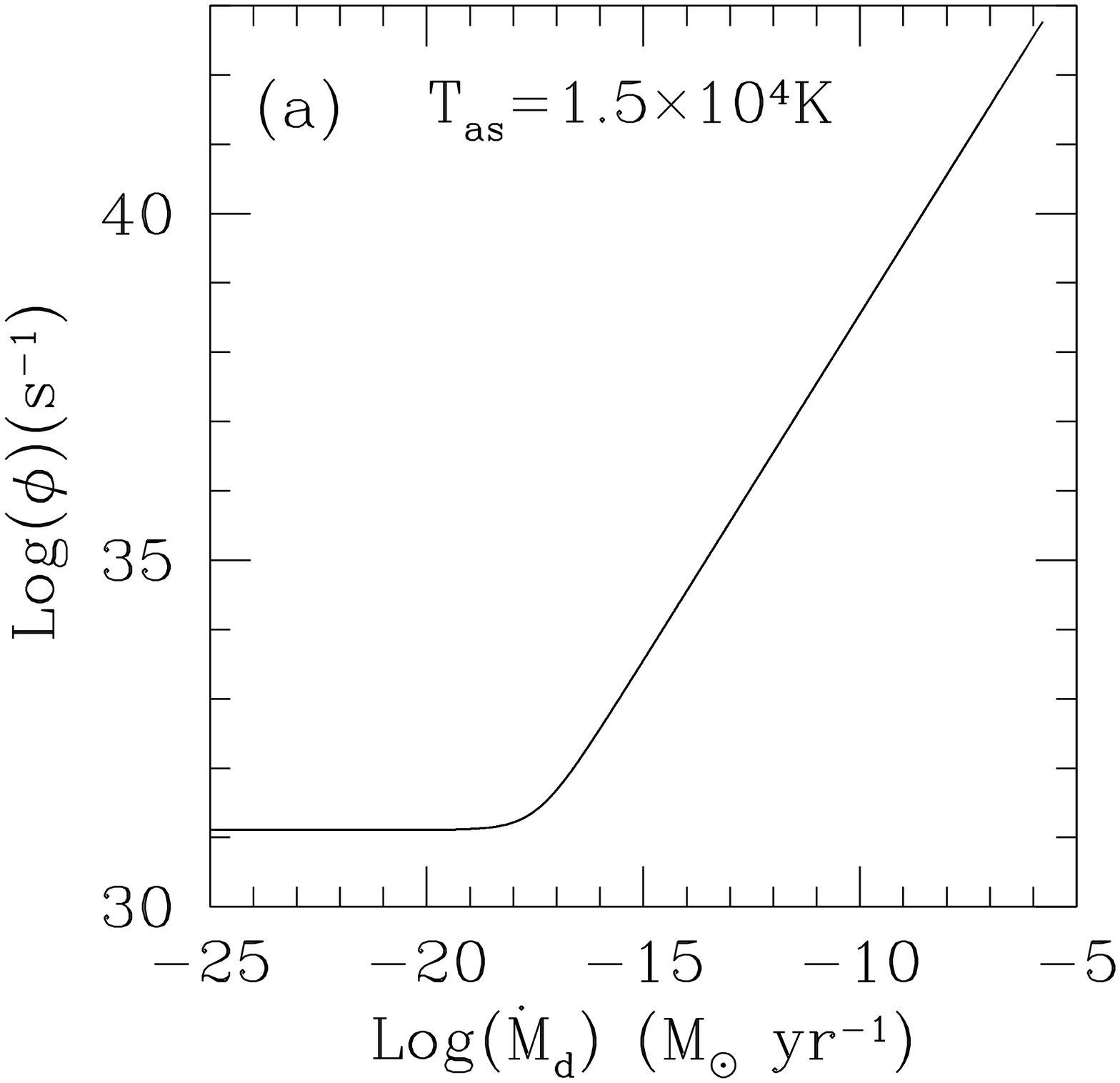}{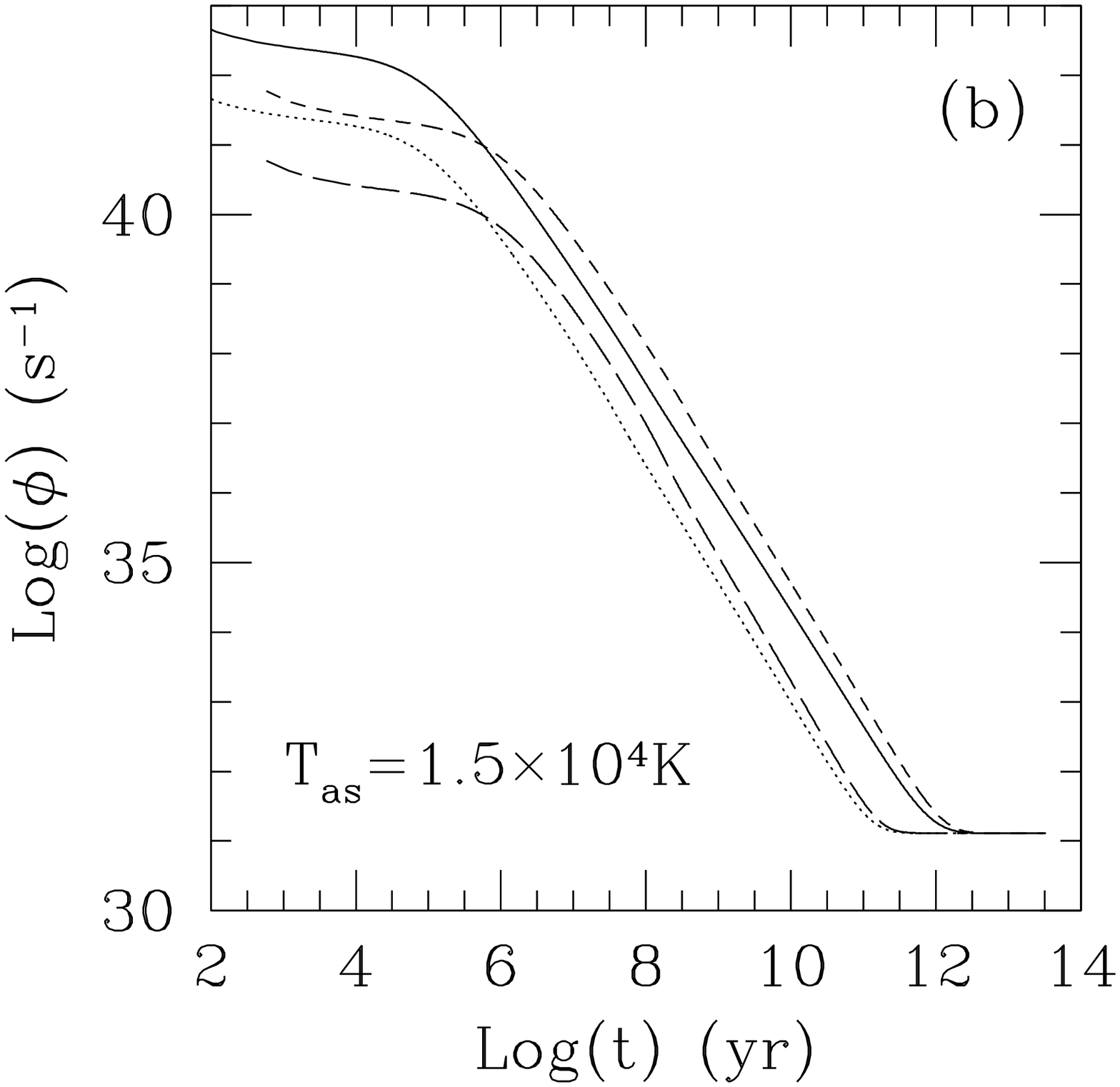} 

\caption[]{Number of ionizing photons, \( \phi  \), as a function
of accretion rate and disk lifetime for different disk parameters
that cover the parameter space from observations. Solid line with
parameters, \( \alpha =10^{-2} \) and \( M_{d}(0)=10^{-1}M_{\sun } \).
Dotted line, \( \alpha =10^{-3} \) and \( M_{d}(0)=10^{-1}M_{\sun } \).
Short-dashed line, \( \alpha =10^{-2} \) and \( M_{d}(0)=10^{-2}M_{\sun
} \).
Long-dashed line, \( \alpha =10^{-3} \) and \( M_{d}(0)=10^{-2}M_{\sun }
\).
The disk is removed by viscous accretion and photoevaporation from
the central source. The ionizing flux approaches the constant value
(\( \phi =1.29\times 10^{31}\textrm{s}^{-1} \)) corresponding to
the quiescent stellar photosphere at late stages of the disk evolution
when
the accretion rates are very small.

\label{central flux} } 

\end{figure}

\citet{uvswitch} used essentially the same model for the
photoevaporation
component of disk dispersal; however, they assumed that the ionizing
flux was constant during the disk evolution. In contrast, we explicitly
model the ultraviolet photons produced in the accretion layer as a
function of time. The total ionizing flux is \( \phi =\phi _{\star
}+\phi _{as} \),
where \( \phi _{\star } \) is the ionizing flux from the stellar
photosphere and \( \phi _{as} \) is the ionizing flux from the accretion
shock. We assume the parameters of typical TTSs under the assumption
that
only the quiescent star produces ultraviolet photons. In fact, these
stars are chromospherically active \citep{chromosphere}, 
providing an additional
source of ionizing photons. However, both the time dependence of the 
chromospheric activity and the rate of ionizing photon production are 
poorly constrained at present and thus we take the limiting case of an 
insignificant chromosphere.  The luminosity is
\( L_{\star }=1L_{\sun } \), the surface temperature is \( T_{\star
}=4000 \)K,
and the accretion shock temperature is \( T_{as}=1.5\times
10^{4}\textrm{K} \) \citep{fuv observations, uv spec}.
We calculate the accretion luminosity using equation \ref{as
luminosity},
and the fraction of ionizing photons from this luminosity with the
accretion shock temperature. For illustration, figure \ref{central
flux}a
plots the corresponding ionizing flux for typical accretion rates
during the disk evolution. The constant ionizing flux (\( \phi
=10^{41}s^{-1} \))
assumed by \citet{uvswitch} is only produced for high accretion rates
(\( \dot{M}\sim 10^{-8}\dot{M}_{\sun }\textrm{yr}^{-1} \)) at early
stages of the disk evolution. We plot in figure \ref{central flux}b
the time dependence of the ionizing flux for several possible disk
scenarios that cover the parameter space of observations. It is clear
that the ionizing flux is not constant, it decreases with the accretion
rate as the disk loses its mass. Even for high viscosities, \( \alpha
\sim 10^{-2}, \)
and initially massive disks, \( M_{d}(0)\sim 10^{-1}M_{\sun } \),
corresponding to high initial accretion rates, the ionizing flux
decreases
to values well below \( \sim 10^{41}\textrm{s}^{-1} \). Thus, it
is essential to compute the ionizing flux self-consistently from the
accretion luminosity when considering photoevaporation from the inner
disk.

Combining photoevaporation with viscous accretion is done numerically.
At each time step, photoevaporation induced mass loss and viscous
accretion induced disk evolution are solved, with the time step chosen
such that the mass removed due to each mechanism is negligible compared
to the instantaneous disk mass. The mass removal rate by
photoevaporation
is determined by equation \ref{central dsigmadt} and the surface
density, since it is only possible to photoevaporate material at
locations
where \( \Sigma >0 \). 

\begin{figure}

\plottwo{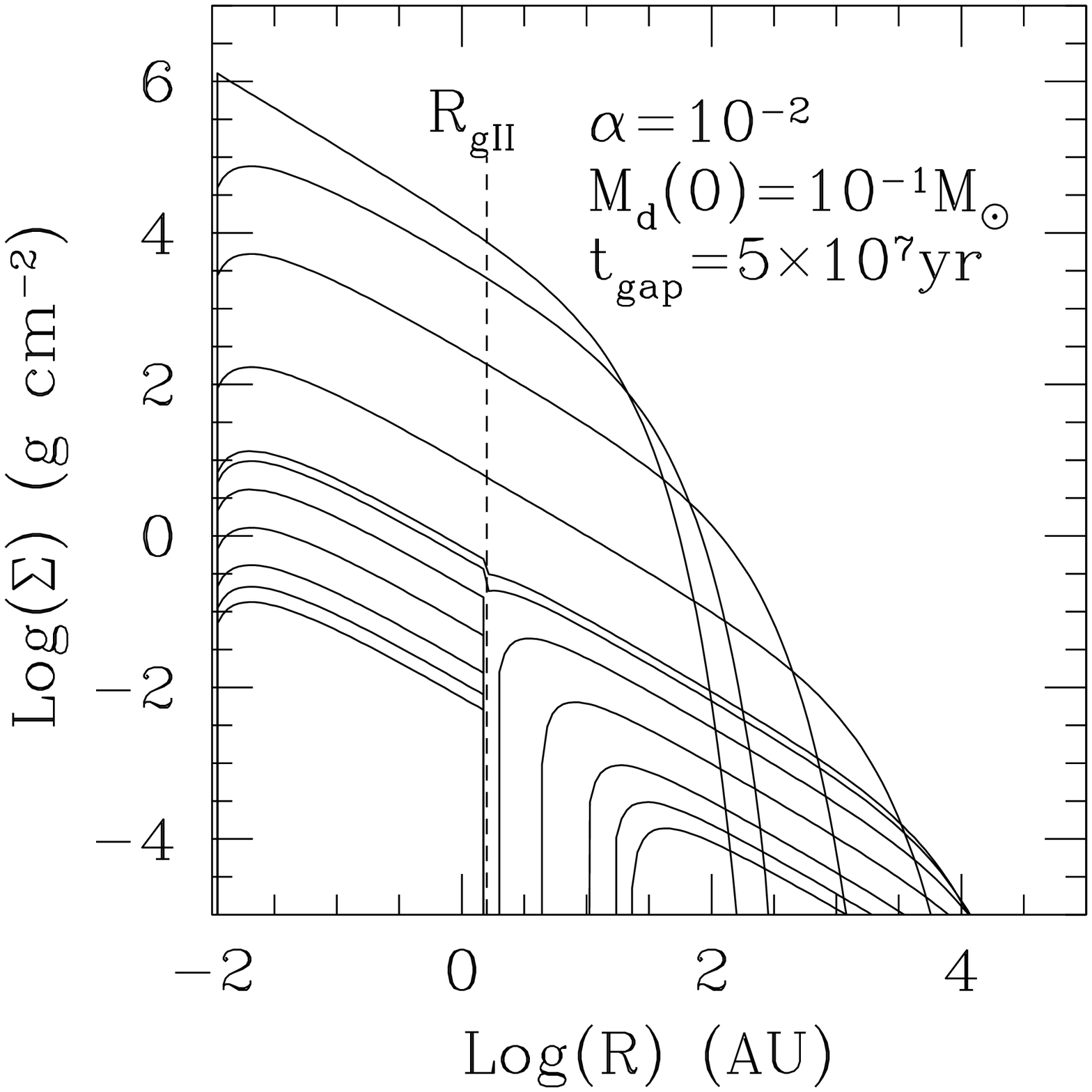}{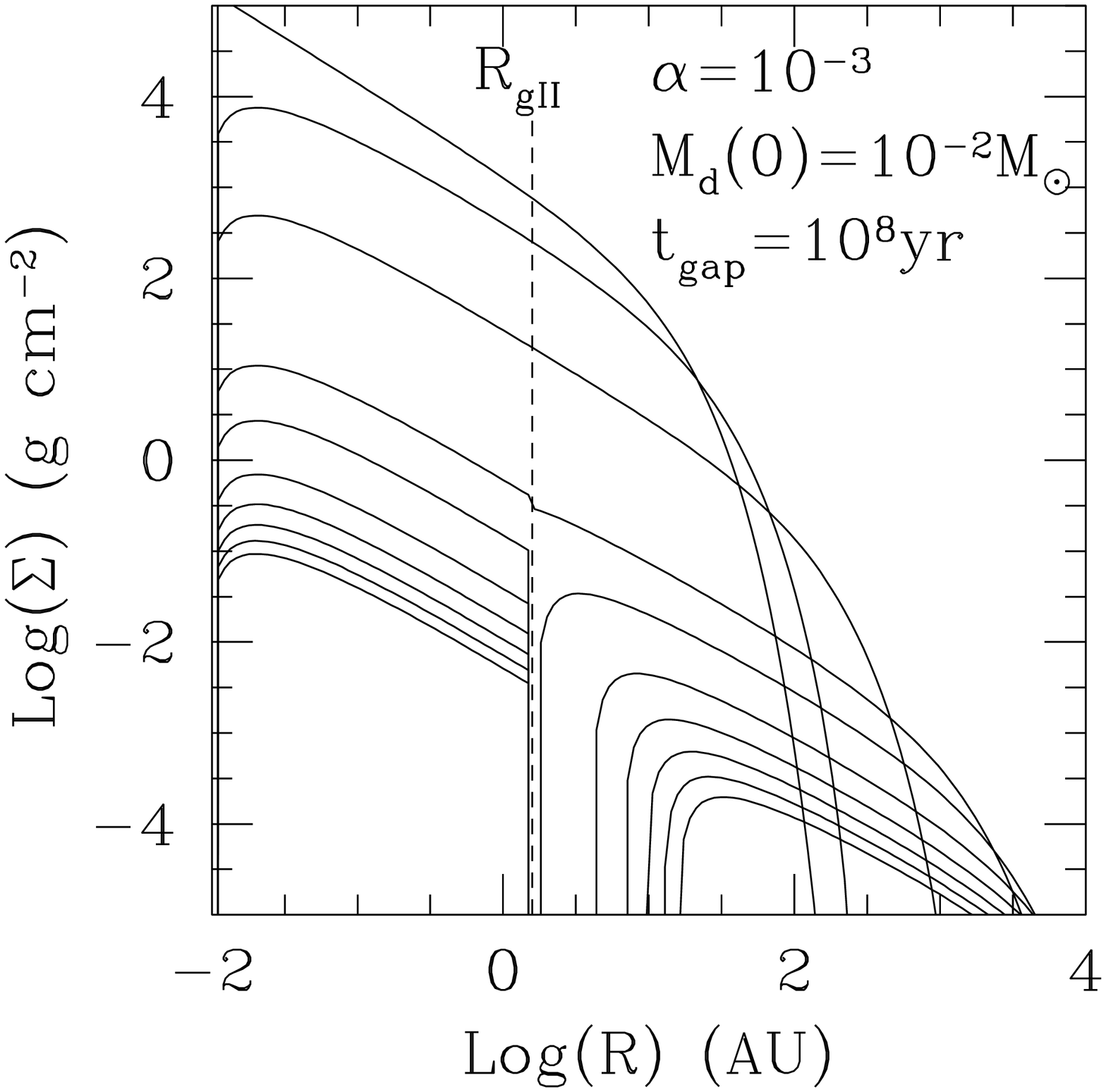} 

\caption[]{Snapshots of the surface density for two representative
models under the influence of viscous diffusion and photoevaporation
from the central source. The short-dashed lines indicate the location
of the gravitational radius, \( r_{gII} \); and \( t_{gap} \) is
the time when gap structures start forming. Left, model with high
viscosity (\( \alpha =10^{-2} \)) and massive initial disk (\(
\textrm{M}_{d}(0)=10^{-1}\textrm{M}_{\sun } \)).
The curves represent \( t=0,10^{5},10^{6},10^{7},5\times 10^{7},\textrm{
}6\times 10^{7},10^{8},2\times 10^{8},4\times 10^{8},6\times
10^{8},\textrm{and }8\times 10^{8}\textrm{yr} \).
A gap structure starts forming at \( t\sim 5\times 10^{7} \)yr, when
the disk mass is \( \sim 3\times 10^{-3}M_{\sun } \). The disk mass
corresponding to the last surface density distribution shown (at \(
t=8\times 10^{8} \)yr)
is \( \sim 10^{-4}\textrm{M}_{\sun } \). Right, model with low viscosity
(\( \alpha =10^{-3} \)) and small initial disk (\(
M_{d}(0)=10^{-2}M_{\sun } \)).
The curves represent \( t=0,10^{6},10^{7},10^{8},2\times 10^{8},\textrm{
}4\times 10^{8},6\times 10^{8},8\times 10^{8},10^{9},\textrm{ and
}1.2\times 10^{9}\textrm{yr} \).
A similar gap structure starts forming at \( t\sim 10^{8} \)yr, when
the disk mass is \( \sim 10^{-3}M_{\sun } \). The disk mass
corresponding
to the last surface density distribution shown (at \( t=1.2\times 10^{9}
\)yr)
is \( \sim 10^{-4}\textrm{M}_{\sun } \). 

\label{central sigma} } 

\end{figure}

\begin{figure}

\plottwo{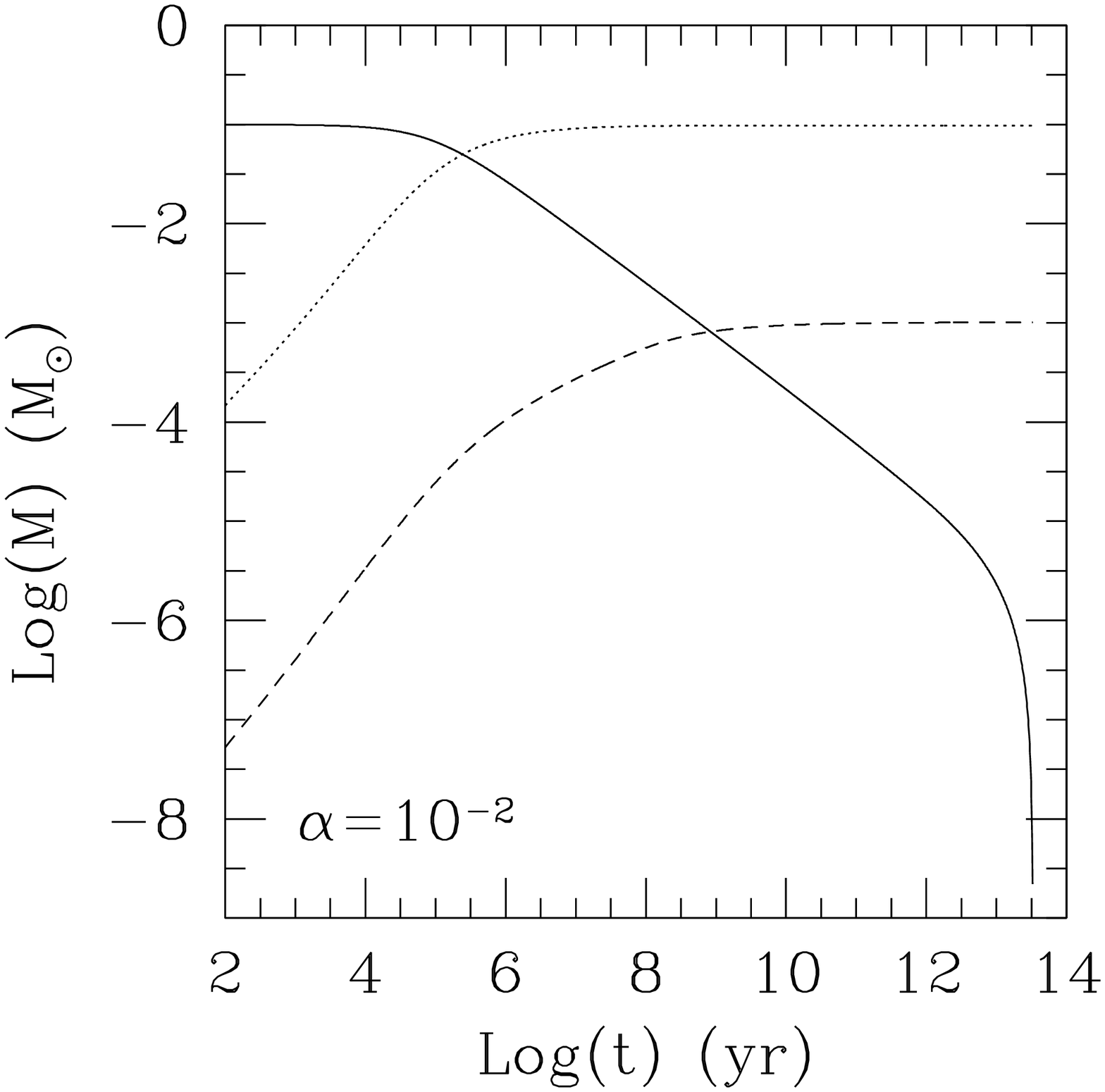}{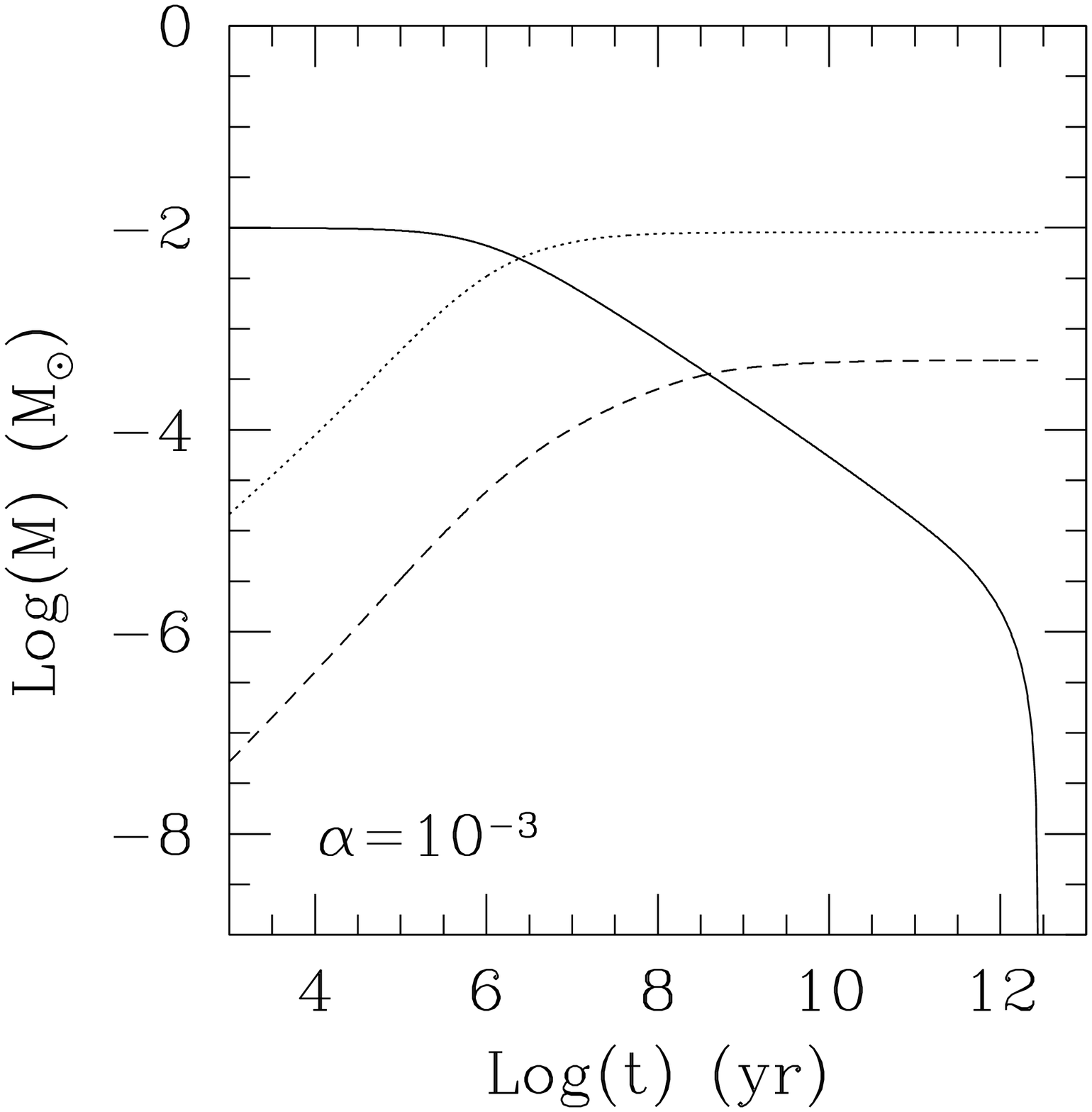}

\caption[]{Disk mass (solid line), disk mass accreted toward the central
star (dotted line), and disk mass removed by photoevaporation from
the central source (short-dashed line) as a function of disk lifetime
for the two representative models. The disk is removed due to viscous
diffusion and photoevaporation from the central source.

\label{central mass} }

\end{figure}

\begin{figure}

\plottwo{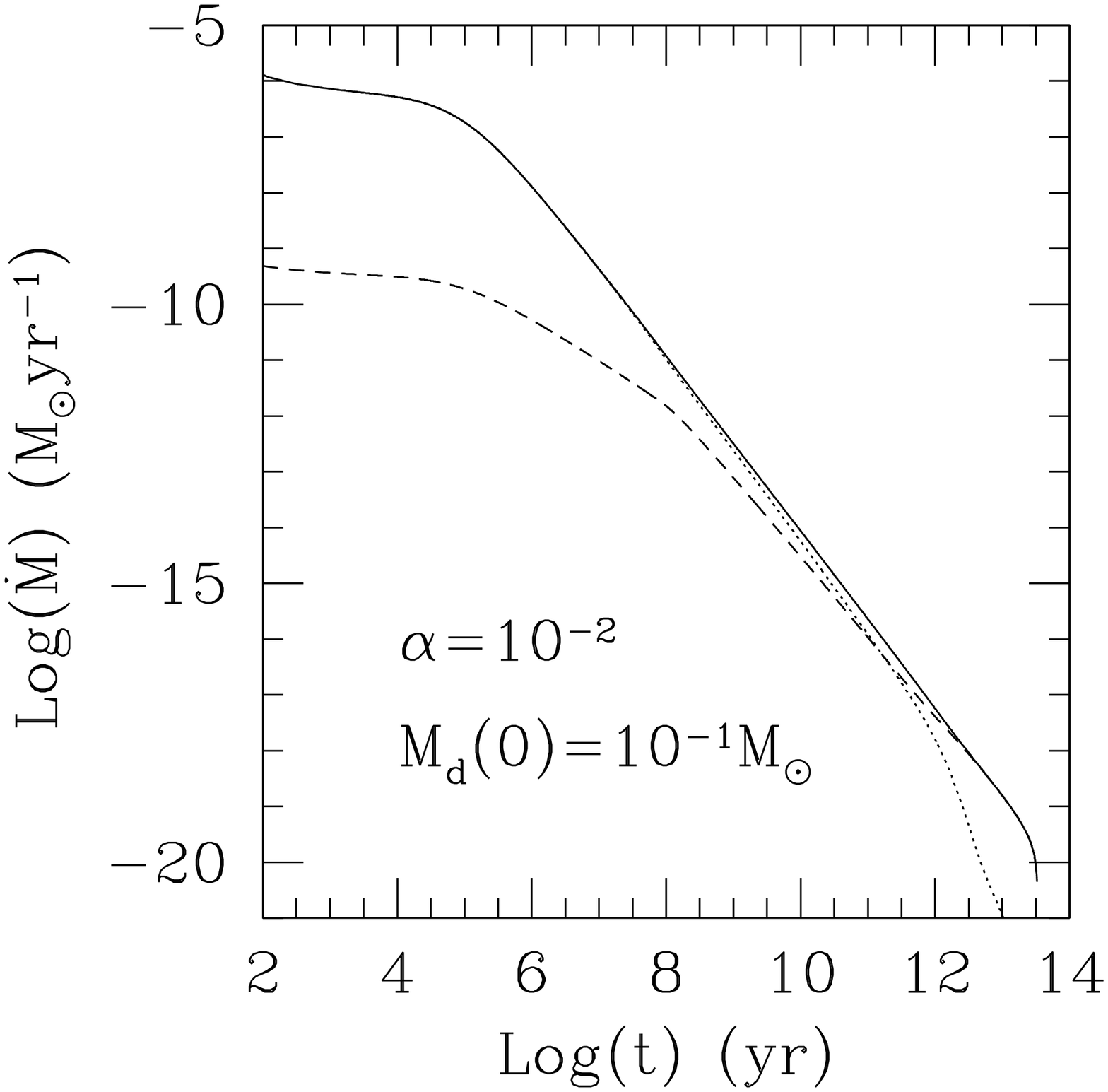}{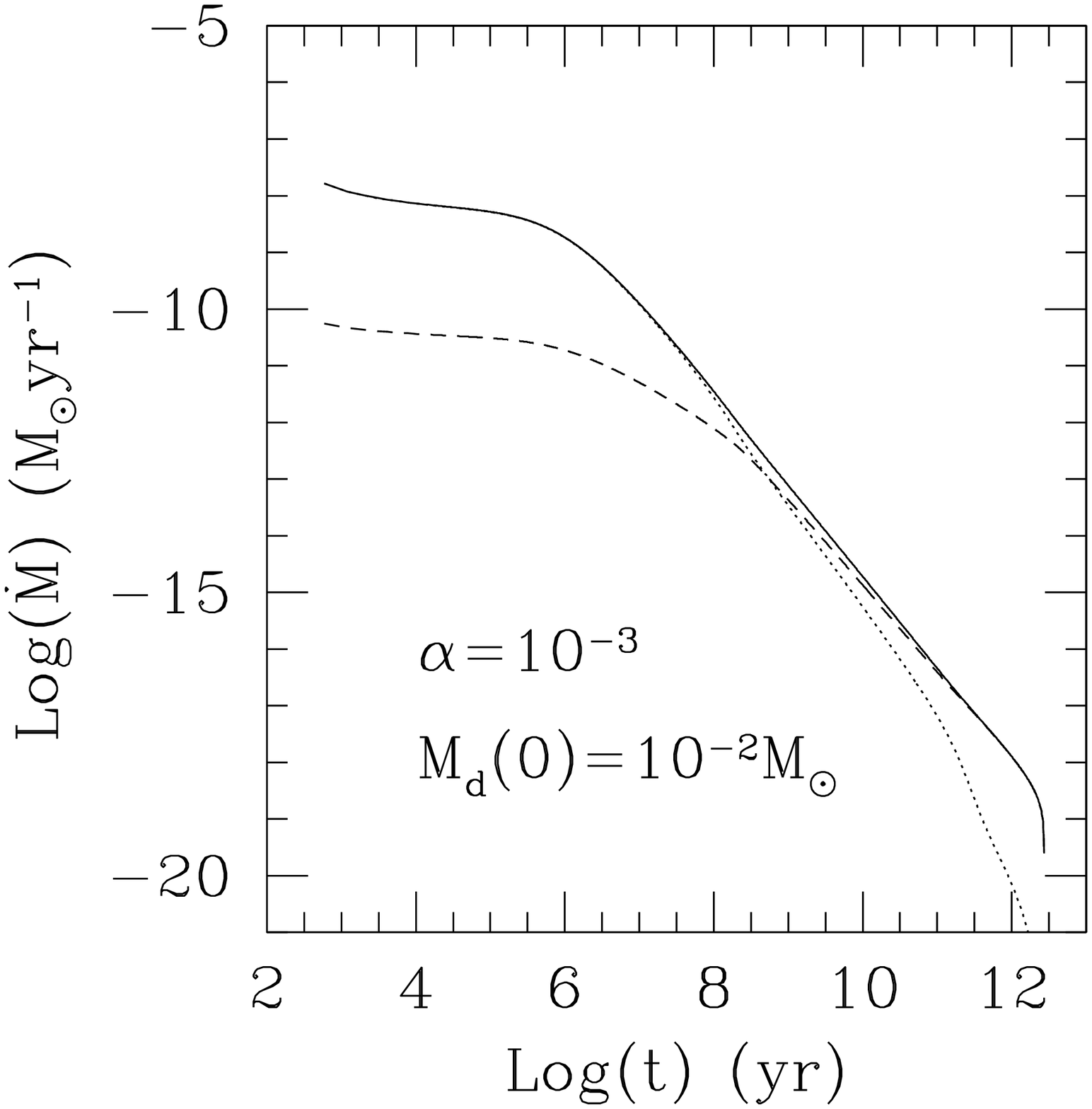}

\caption[]{Total mass removal rate (solid line), mass removal rate
due to accretion (dotted line), and mass removal rate due to
photoevaporation
by the central source (short-dashed line) vs. disk lifetime for the
two representative models. 

\label{central mdot} } 

\end{figure}

Figure \ref{central sigma} shows snapshots of the surface density
distribution for two representative cases: a model with high viscosity
(\( \alpha =10^{-2} \)) and a massive initial disk (\(
M_{d}(0)=10^{-1}M_{\sun } \)),
and a model with low viscosity (\( \alpha =10^{-3} \)) and a small
initial disk (\( M_{d}(0)=10^{-2}M_{\sun } \)). In both cases a gap
structure forms during the later stages of the disk evolution. Disk
material is accreted toward the central star and photoevaporation
from the central source removes material outside the gravitational
radius. Figure \ref{central mass} shows the remaining disk mass,
the disk mass accreted toward the central star, and the disk mass
removed by photoevaporation. Most of the initial disk mass is accreted
toward the central star in both cases. Photoevaporation removes all
the material in the vicinity of the gravitational radius when the
surface density at this radius is low (\( \sim 10^{-1}\textrm{g
cm}^{-2}) \),
and divides the disk into an inner and an outer annulus. The subsequent
evolution is dominated by two counteracting effects; viscous diffusion
attempts to spread both annuli and remove the gap structure while
photoevaporation removes material predominantly at the gravitational
radius and reopens the gap. The outcome of the combination of the
two mechanisms is an efficient mass removal from the disk. Disk material
at the outer edge of the inner annulus is viscously spread beyond
the gravitational radius, where it is removed by photoevaporation.
\citet{uvswitch} showed that it is possible to quickly remove the
inner disk; however, we show that the inner disk is removed at the
same rate as the outer disk. This difference arises because
\citet{uvswitch}
assume a constant ionizing flux while we calculate the ionizing flux
from the accretion luminosity. In our model, the ionizing flux from
the accretion shock decreases (see figure \ref{central flux}b) as
the accretion rate decreases (see figure \ref{central mdot}) until
it reaches the constant value (\( 1.29\times 10^{31}\textrm{s}^{-1} \))
corresponding to the quiescent stellar photosphere. Therefore, removing
the
inner disk and maintaining a high ionizing flux (\( \sim
10^{41}\textrm{s}^{-1} \))
is not self-consistent. It is not possible to remove the inner disk
faster than the outer disk even for high accretion shock temperatures
(\( T_{as}\sim 3\times 10^{4} \)K) that correspond to stronger ionizing
fluxes. The strong dominance of viscous diffusion over photoevaporation
(see figures \ref{central mass} and \ref{central mdot}) produce 
unrealistically long disk lifetimes (\( \sim 10^{12}-10^{13}\textrm{yr}
\);
see figure \ref{central mass}) unless the stellar ionizing flux is
extremely enhanced through an active chromosphere (Figure
\ref{central flux}b can be used to estimate the time at which this
model would break down if the chromospheric activity produced a
significant
ionizing flux $\phi_{ch}$. For example, if $\phi_{ch} =
10^{35}\textrm{s}^{-1}$
for more than $\sim 10^{9}\textrm{yr}$ then the calculated models would
begin to diverge from reality.) It is not possible to quickly remove
the inner annulus due to the vanishing ionizing flux. However, as
the outer annulus spreads to distances far from the gravitational
radius where photoevaporation becomes inefficient, other removal
mechanisms 
such as stellar encounters become important and will limit the disk
lifetime.

\subsection{EUV photoevaporation by external stars}

Stellar disks are also dispersed by external stars and this is a likely
situation for the disks surrounded by ionization fronts in the Trapezium
Cluster \citep{orion observations}. \citet{j98} found models for EUV
dominated external photoevaporation based on observations of the
proplyds
in the Orion Nebula. In this scenario the disk is heated by UV photons
from the nearby O stars. Since the inner annuli have a small surface
area compared to the outer annuli, their contribution to the process
is very small. Most of the material is removed at the disk edge, \(
R_{d} \),
and photoevaporation from the disk can be approximated by
photoevaporation
from a sphere with radius \( R_{d} \). The EUV photons control the
flux close to the central star; heating the gas to $
T_{\textrm{\ion{H}{2}}}\sim 10^{4}\textrm{K} $
and creating an ionization front. Following \citet{j98}, the mass
loss rate for EUV photoevaporation can be approximated by

\begin{equation}\label{centralEUV dmdt} \dot{M}_{d}^{EUV}=\cases{
7\times 10^{-12}\textrm{ M}_{\sun }\textrm{ yr}^{-1}\left( \frac{\Phi
_{i}}{10^{49}\textrm{s}^{-1}}\right) ^{1/2}\left(
\frac{1\textrm{pc}}{d}\right) \left( \frac{R_{d}}{1\textrm{ AU}}\right)
^{3/2}, &if $R_{d}>\beta R_{gII}$;\cr 0, &otherwise; \cr }
\end{equation}
where \( \phi _{i} \) is the ionization rate of the external star
and \( d \) is the distance to the external star. Most of the mass
is evaporated from the outer disk annuli because they have the largest
surface area. We assume the Trapezium Cluster conditions, where most
of the flux, \( \phi _{i}\sim 10^{49}\textrm{s}^{-1} \) is from \(
\theta ^{1}\textrm{ Ori C} \).
The EUV photons dominate the flux for the proplyds orbiting at a
distance
\( d\la 0.03 \)pc from the external star \citep{j98, sh99}, and we
assume
\( d=0.02 \)pc in the following models. The surface density evaporation
rate is the mass loss rate divided by the effective area of the disk:

\begin{equation} \label{centralEUV dsigmadt} \dot{\Sigma }^{EUV}(R)=
\cases { \frac{\dot{M}_{d}^{EUV}}{\pi (R_{d}^{2}-\beta
^{2}R_{gII}^{2})}, &if $R>\beta R_{gII}$; \cr 0, &otherwise. \cr }
\end{equation}

\begin{figure}

\plottwo{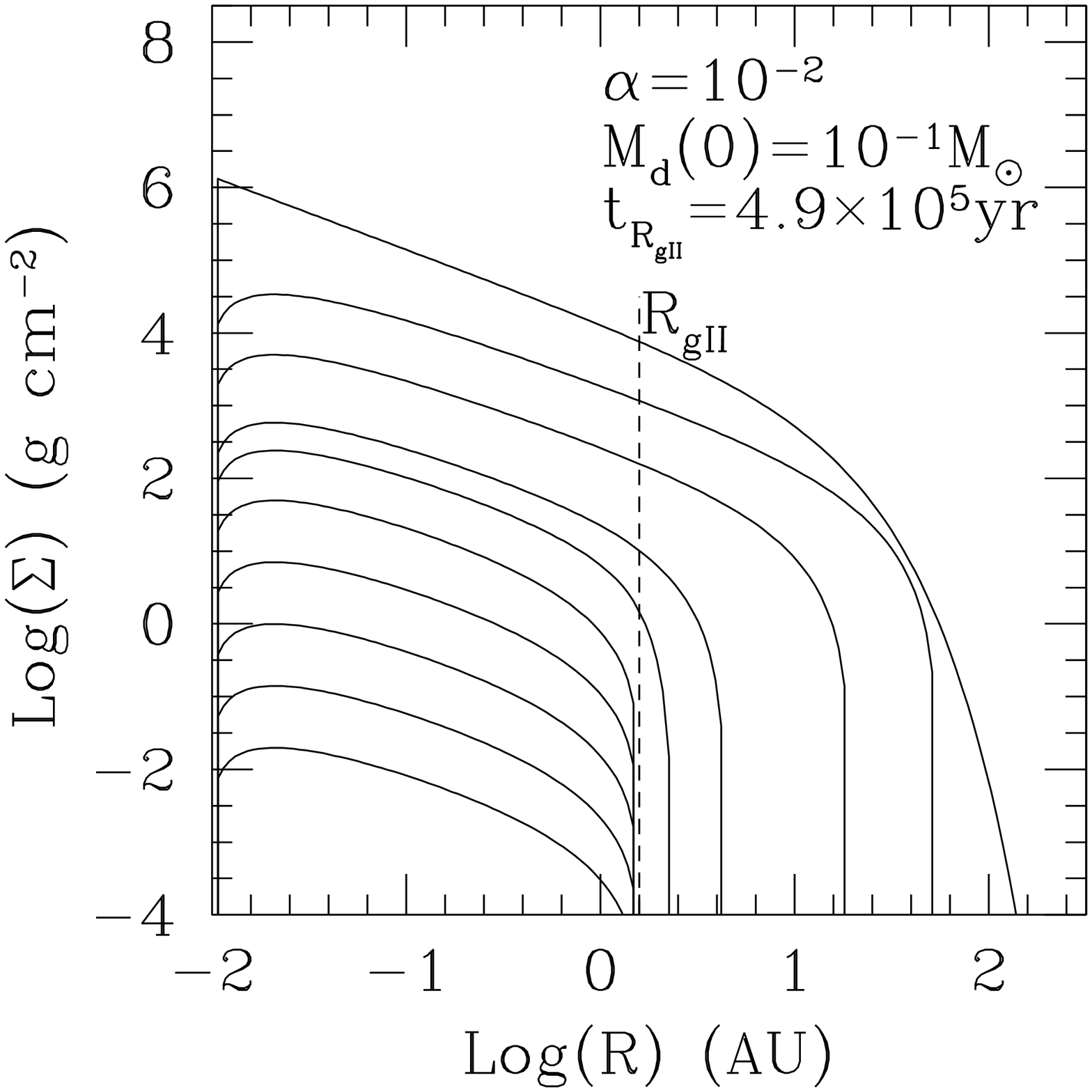}{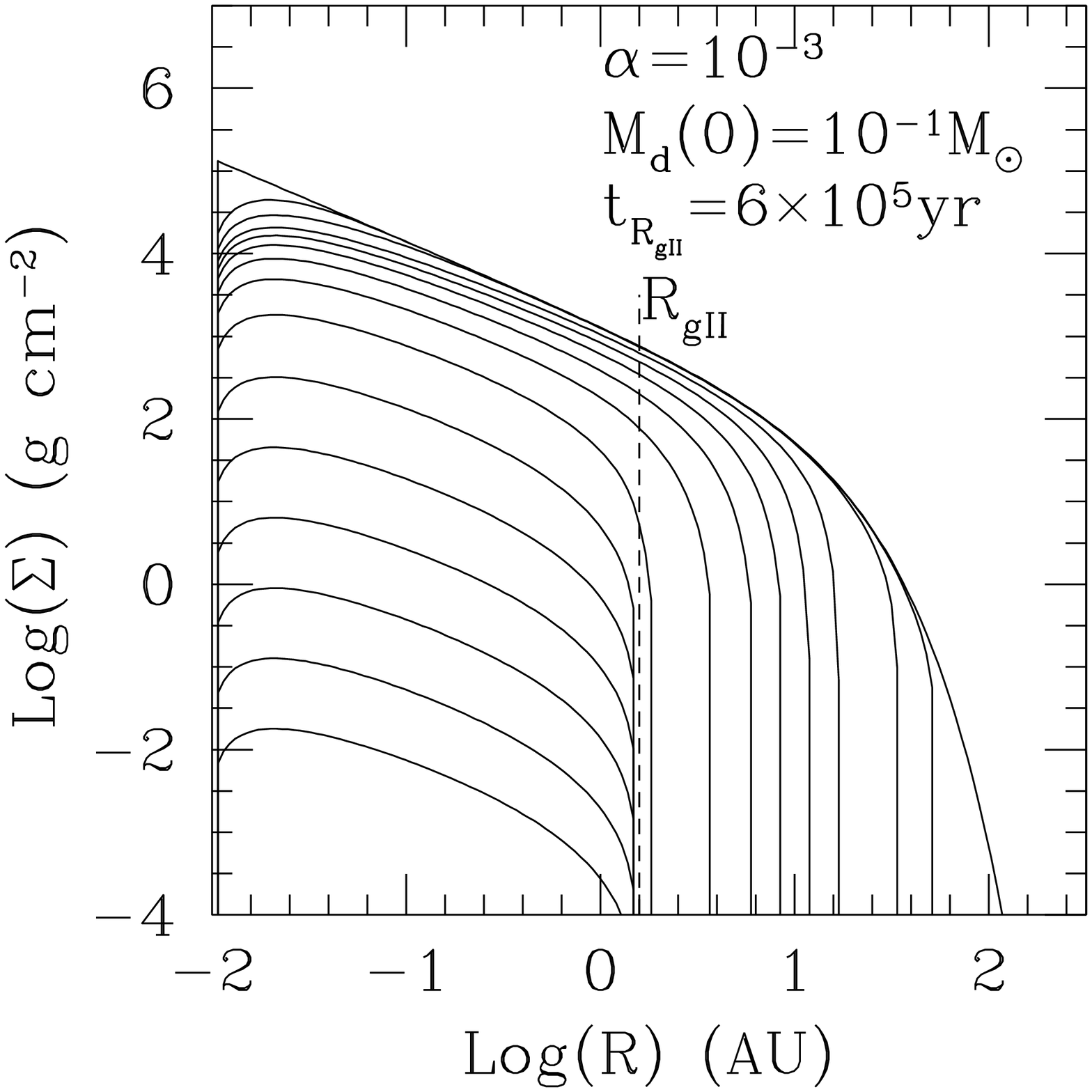} 

\caption[]{Snapshots of the surface density for two representative
models considering viscous diffusion, photoevaporation from the central
source, and EUV photoevaporation by external stars. The dashed lines
indicate the location of the EUV gravitational radius, \( r_{gII} \);
and \( t_{gap} \) is the time when gap structures start forming.
Left, model with high viscosity (\( \alpha =10^{-2} \)) and massive
initial disk (\( \textrm{M}_{d}(0)=10^{-1}\textrm{M}_{\sun } \)).
The curves represent \( t=0,2\times 10^{5},4\times 10^{5},4.7\times
10^{5},4.\textrm{ }8\times 10^{5},4.9\times 10^{5},5\times
10^{5},5.1\times 10^{5},5.2\times 10^{5},\textrm{and }5.3\times
10^{5}\textrm{yr} \).
The disk edge reaches the EUV gravitational radius at \( t_{R_{gII}}\sim
4.9\times 10^{5}yr \),
when the disk mass is \( \sim 10^{-6}\textrm{M}_{\sun } \). The disk
mass corresponding to the last surface density distribution shown
(at \( t=5.3\times 10^{5} \)yr) is \( \sim 10^{-9}\textrm{M}_{\sun } \).
Right, model with low viscosity (\( \alpha =10^{-3} \)) and small
initial disk (\( M_{d}(0)=10^{-2}M_{\sun } \)). The curves represent
\( t=0,10^{3},10^{4},10^{5},2\times 10^{5},\textrm{ }3\times
10^{5},4\times 10^{5},5\times 10^{5},6\times 10^{5},7\times
10^{5},8\times 10^{5},9\times 10^{5},10^{6},\textrm{ }1.1\times
10^{6},\textrm{ and }1.2\times 10^{6}\textrm{yr} \).
The disk edge reaches the EUV gravitational radius at \( t_{R_{gII}}\sim
6\times 10^{5} \)yr,
when the disk mass is \( \sim 10^{-5}\textrm{M}_{\sun } \). The disk
mass corresponding to the last surface density distribution shown
(at \( t=1.2\times 10^{6} \)yr) is \( \sim 10^{-9}\textrm{M}_{\sun } \).

\label{centralEUV sigma} } 

\end{figure}

\begin{figure}

\plottwo{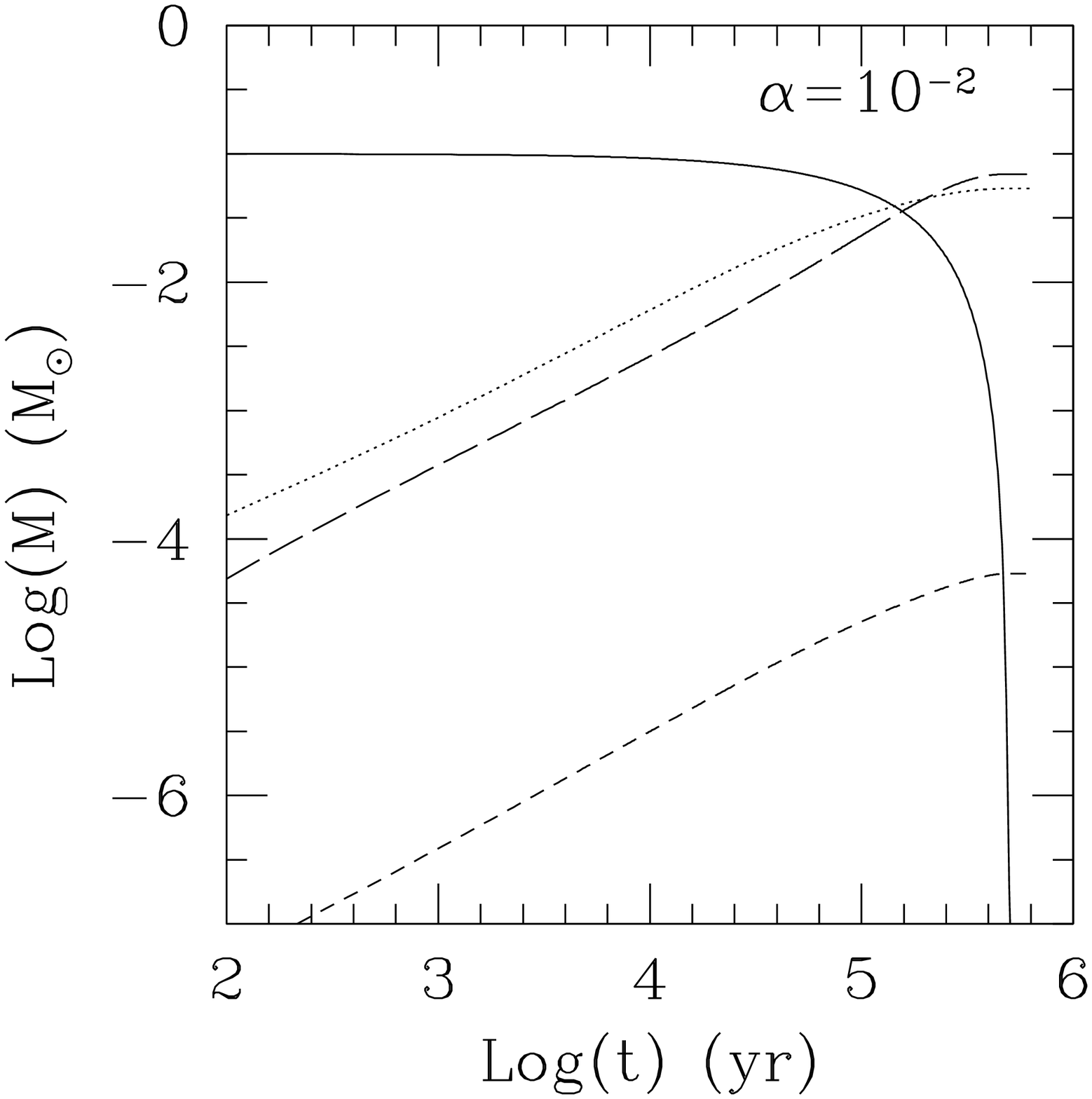}{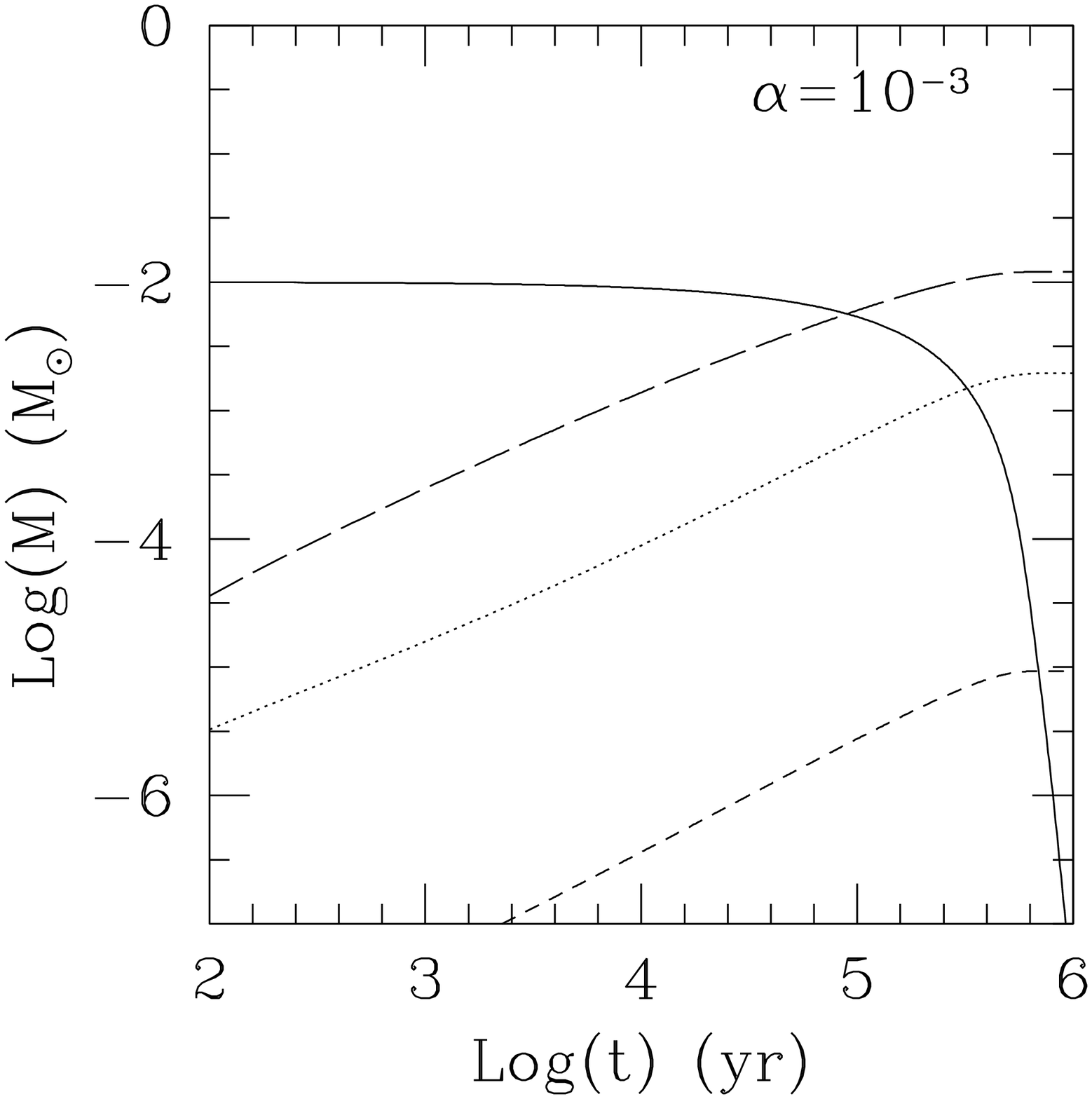}

\caption[]{Disk mass (solid line), disk mass accreted toward the central
star (dotted line), disk mass removed by photoevaporation from the
central source (short-dashed line), and disk mass removed by EUV
photoevaporation
by external stars (long-dashed line) vs. disk lifetime for the two
representative models. 

\label{centralEUV mass} } 

\end{figure}

\begin{figure}

\plottwo{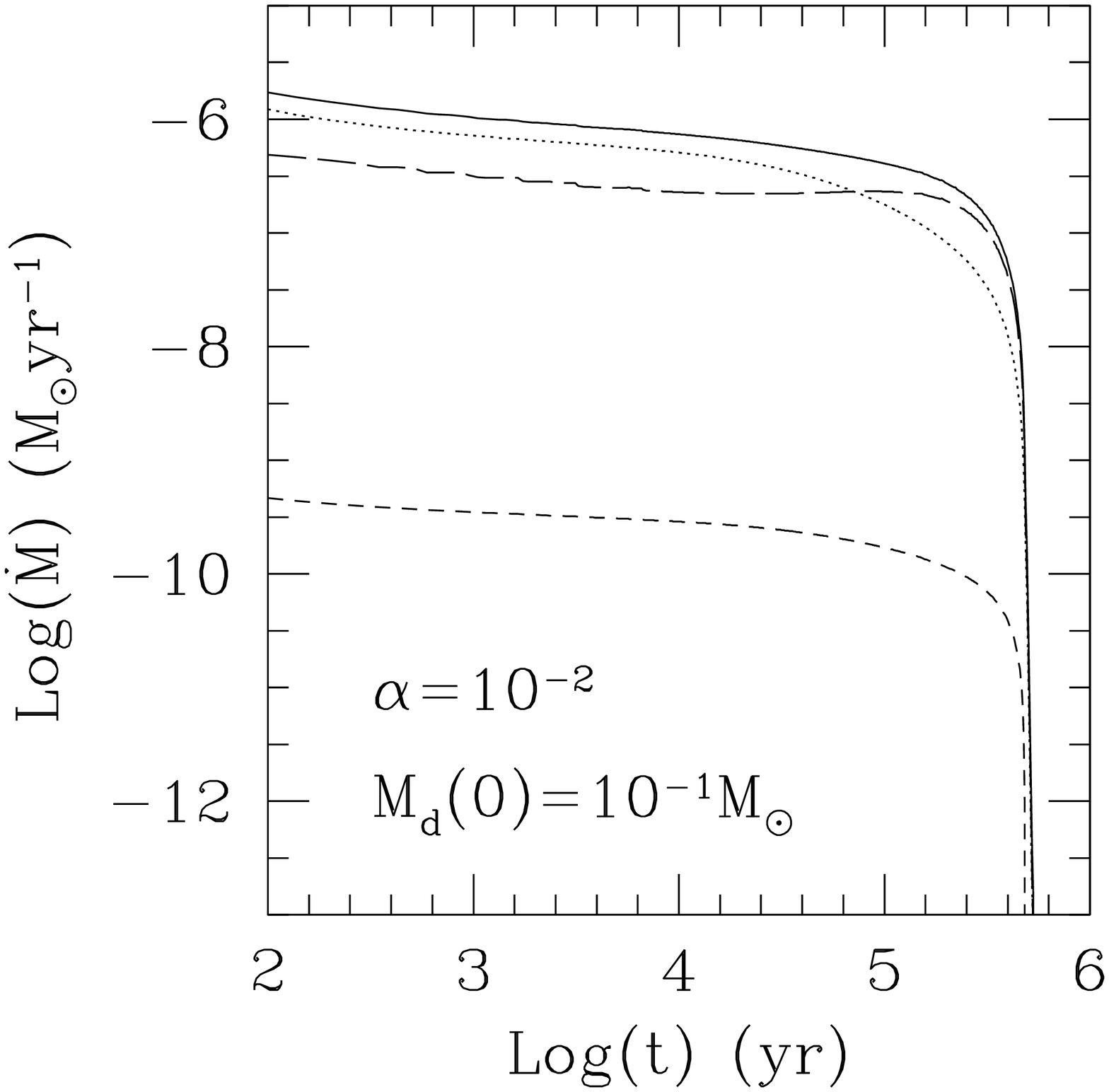}{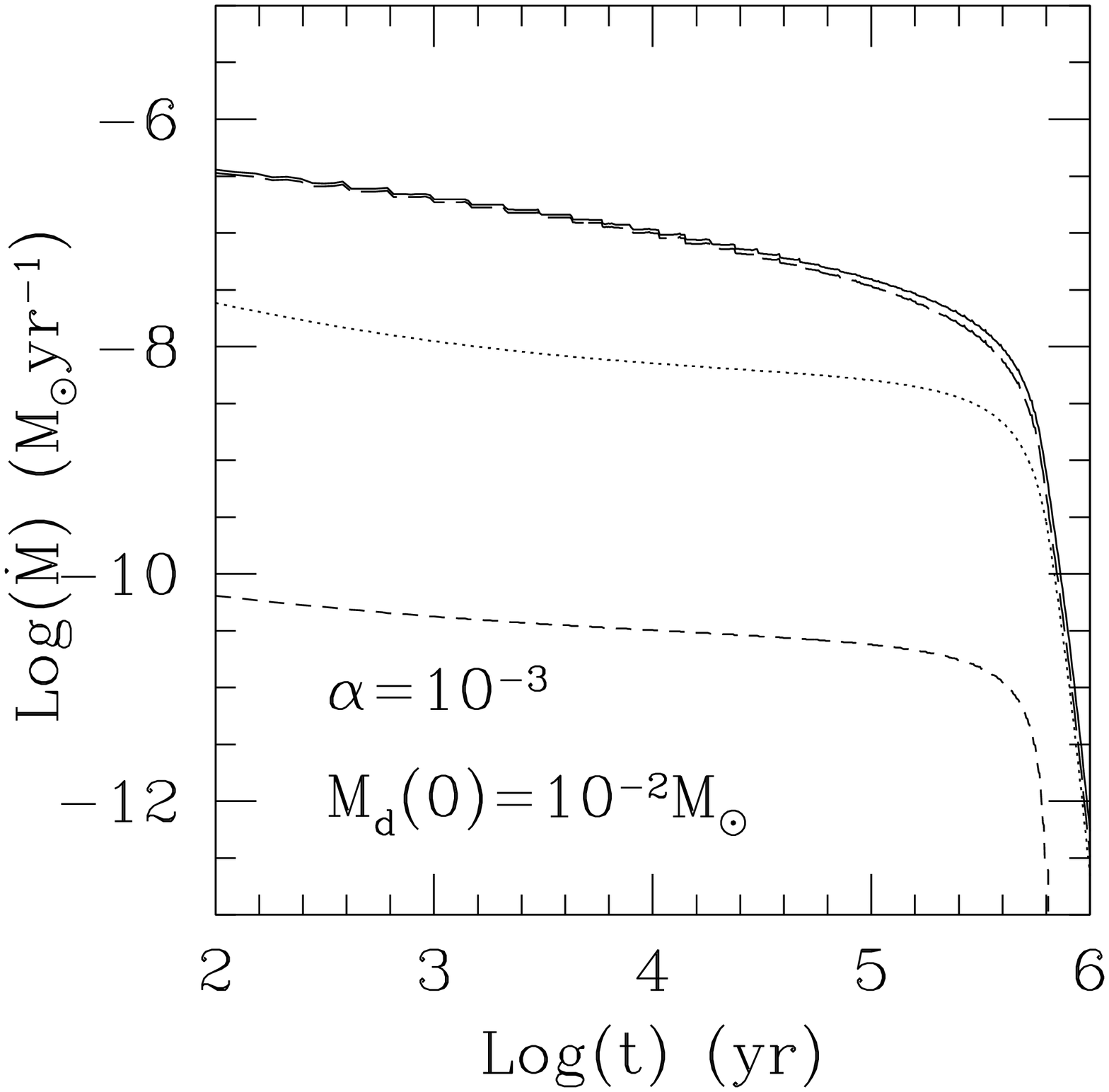}

\caption[]{Total mass removal rate (solid line), mass removal rate
due to accretion (dotted line), mass removal rate due to
photoevaporation
by the central source (short-dashed line), and disk mass removed by EUV 
photoevaporation
by external stars (long-dashed line) as a function of disk lifetime
for the two representative models. The disk is removed by viscous
diffusion, photoevaporation from the central source, and EUV
photoevaporation
by external stars. 

\label{centralEUV mdot} } 

\end{figure}

The disk is removed due to accretion toward the central star,
photoevaporation
from the central source, and EUV photoevaporation by the external
stars. The disk evolution is strongly affected by the external radiation
field. Photoevaporation by the external star dominates the ionizing
flux at large radii where material is efficiently evaporated. Hence,
photoevaporation by the central star is a minor disk removal mechanism.
We show the results for two models with the same parameters as before
(see section 3.1). Figure \ref{centralEUV sigma} shows snapshots
of the surface density distribution. Viscous diffusion spreads the
disk in both directions and photoevaporation from the external star
removes the outer parts; as a result, the disk size remains constant
(at the gravitational radius) until just before the disk is completely
removed. The disk truncation is in good agreement with the two
dimensional 
simulations of \citet{richling}. However, their simulation stops at 
$t\sim10^{4}\textrm{yr}$, before the disk is completely removed. 
There is no formation of gap structures due to the strong dominance of
EUV 
photoevaporation from the external stars over photoevaporation from the
central source. 
The disk is quickly removed as viscous accretion, photoevaporation by
the central star, 
and EUV photoevaporation by the external star work in resonance. The
strong dominance of 
photoevaporation by the external star is also illustrated in the total
mass removed
by photoevaporation (see figure \ref{centralEUV mass}) and the mass
removal rate (see figure \ref{centralEUV mdot}). Photoevaporation
by the external stars removes most of the disk mass and changes
dramatically
the disk evolution, in particular, it is possible to completely remove
the disk in \( \sim 10^{6} \)yr.

\subsection{FUV photoevaporation by external stars}

While the EUV photons create an ionization front and heat the gas
to the temperature, $ T_{\textrm{\ion{H}{2}}}\sim 10^{4}\textrm{K} $,
the FUV photons penetrate this ionized region and heat a neutral
hydrogen
layer to the temperature, $ T_{\textrm{\ion{H}{1}}}\sim 10^{3}\textrm{K}
$.
At large distances from the external star the FUV photons dominate
and the neutral layer launches a supersonic flow. \citet{j98} also
found models for FUV dominated external photoevaporation based on
observations of the proplyds in the Orion Nebula. The mass loss rate
by photoevaporation can be approximated by

\begin{equation}\label{centralFUV dmdt} \dot{M}^{FUV}_{d}=\cases{
2\times10^{-9}\textrm{ M}_{\sun }\textrm{ yr}^{-1}\left(
\frac{N_{D}}{5\times 10^{21}\textrm{cm}^{-2}}\right) \left(
\frac{R_{d}}{1\textrm{ AU}}\right), &if $R_{d}>R_{gI}$; \cr 0,
&otherwise; \cr } \end{equation}
where \( N_{D} \) is the column density from the ionization front
to the disk surface, and \( R_{gI} \) is the gravitational radius
for the neutral layer. In the following models it is assumed that
\( N_{D}\sim 5\times 10^{21}\textrm{cm}^{-2} \) based on the numerical
results of \citet{pdr98}. The surface density photoevaporation rate
for the FUV dominated flow is the mass loss rate divided by the
effective
area of the disk:

\begin{equation}\label{centralFUV dsigmadt}
\dot{\Sigma}^{FUV}(R)=\cases{ \frac{\dot{M}_{d}^{FUV}}{\pi
(R_{d}^{2}-\beta ^{2}R_{gII}^{2})}, &if $R>\beta R_{gI}$; \cr 0,
&otherwise. \cr } \end{equation}

The disk mass is removed due to viscous accretion toward the central
star, photoevaporation from the central source, and FUV photoevaporation
by the external stars. For the Trapezium Cluster, the FUV photons
dominate the flux for the proplyds orbiting at a distance \( d\gg 0.03
\)pc
from the external stars \citep{j98, sh99}.

\begin{figure}

\plottwo{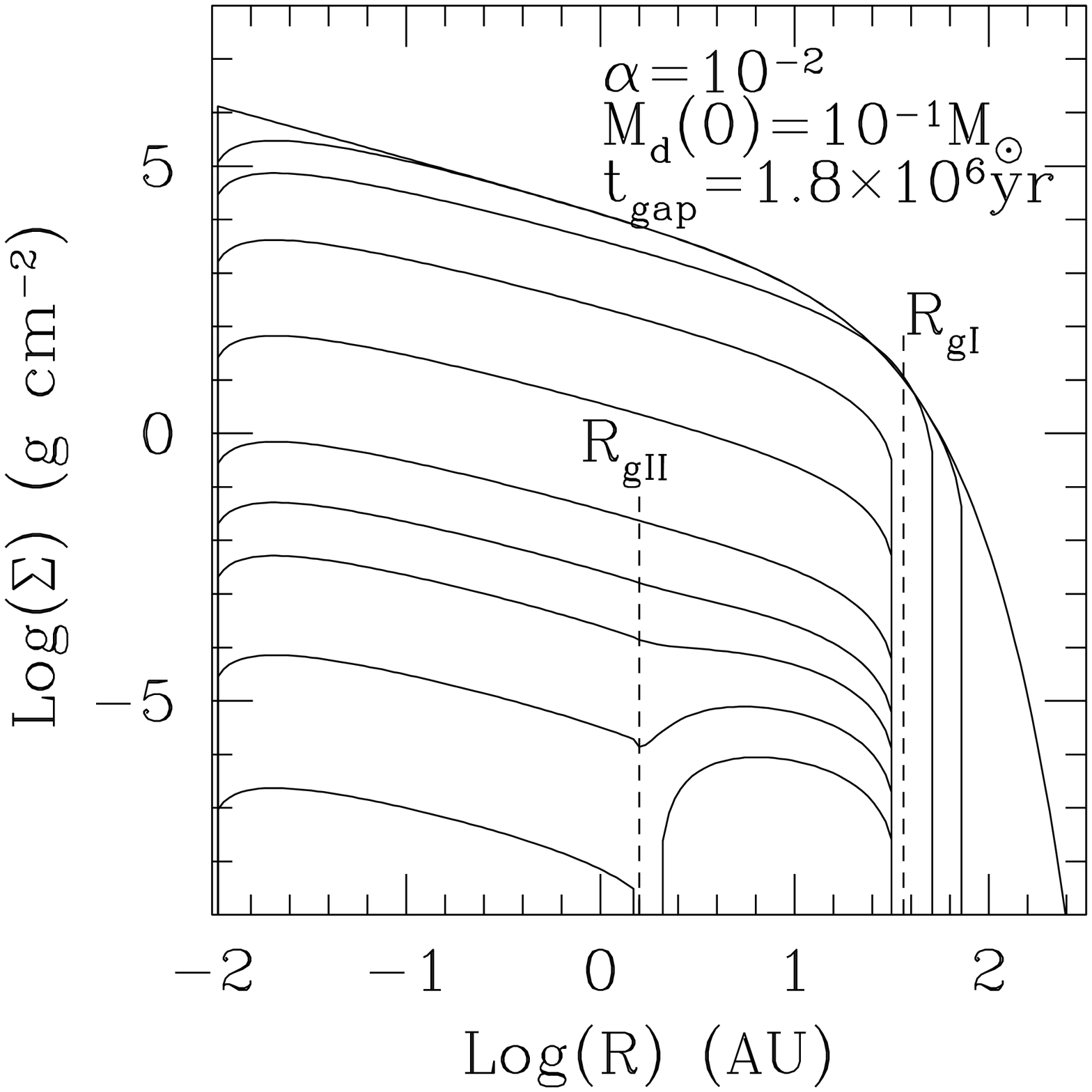}{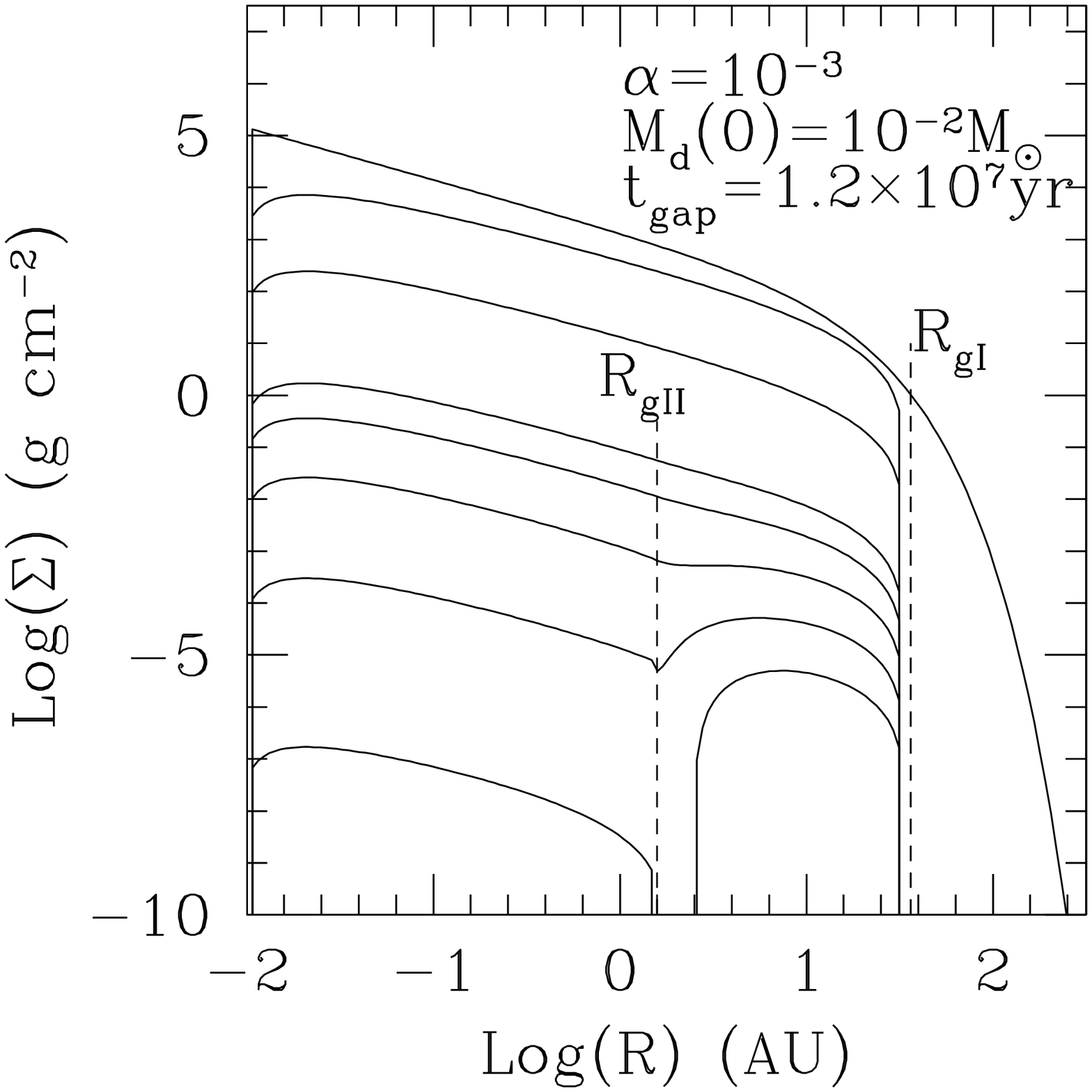} 

\caption[]{Snapshots of the surface density for two representative
models with viscous diffusion, photoevaporation from the central source,
and EUV photoevaporation by external stars. The dashed lines indicate
the location of the EUV gravitational radius, \( r_{gII} \), the
FUV gravitational radius, \( r_{gI} \). A gap structure forms at
\( t\sim t_{gap} \). Left, model with high viscosity (\( \alpha =10^{-2}
\))
and massive initial disk (\( M_{d}(0)=10^{-1}M_{\sun } \)). The curves
represent \( t=0,10^{3},10^{5},5\times 10^{5},10^{6},1.5\times
10^{6},1.7\times 10^{6},1.8\times 10^{6},\textrm{ }1.9\times
10^{6},\textrm{ and }2\times 10^{6}\textrm{yr} \).
The disk edge reaches the FUV gravitational radius, \( R_{gI} \),
at \( t\sim 10^{5}yr \), when the disk mass is \( \sim 3\times
10^{-2}M_{\sun } \).
A gap structure forms at \( t_{gap}\sim 1.8\times 10^{6} \)yr, when
the disk is almost completely removed (\( M_{d}\sim 10^{-8}M_{\sun }
\)).
The disk mass corresponding to the last surface density distribution
shown (at \( t=2\times 10^{6} \)yr) is \( \sim 10^{-10}\textrm{M}_{\sun
} \).
Right, model with low viscosity (\( \alpha =10^{-3} \)) and small
initial disk (\( M_{d}(0)=10^{-2}M_{\sun } \)). The curves represent
\( t=0,10^{6},5\times 10^{6},10^{7},1.1\times 10^{7},1.2\times
10^{7},1.3\times 10^{7},\textrm{ and }1.4\times 10^{7}\textrm{yr} \).
The disk edge is reduced to the FUV gravitational radius, \( R_{gI} \),
at \( t\sim 10^{6} \)yr, when the disk mass is \( \sim 4\times
10^{-3}M_{\sun } \).
A gap structure starts forming at \( t_{gap}\sim 1.2\times 10^{7} \)yr,
when the disk mass is reduced to \( \sim 10^{-7}M_{\sun } \). The
disk mass corresponding to the last surface density distribution shown
(at \( t=1.4\times 10^{7} \)yr) is \( \sim 10^{-9}\textrm{M}_{\sun } \).

\label{centralFUV sigma} } 

\end{figure}

\begin{figure}

\plottwo{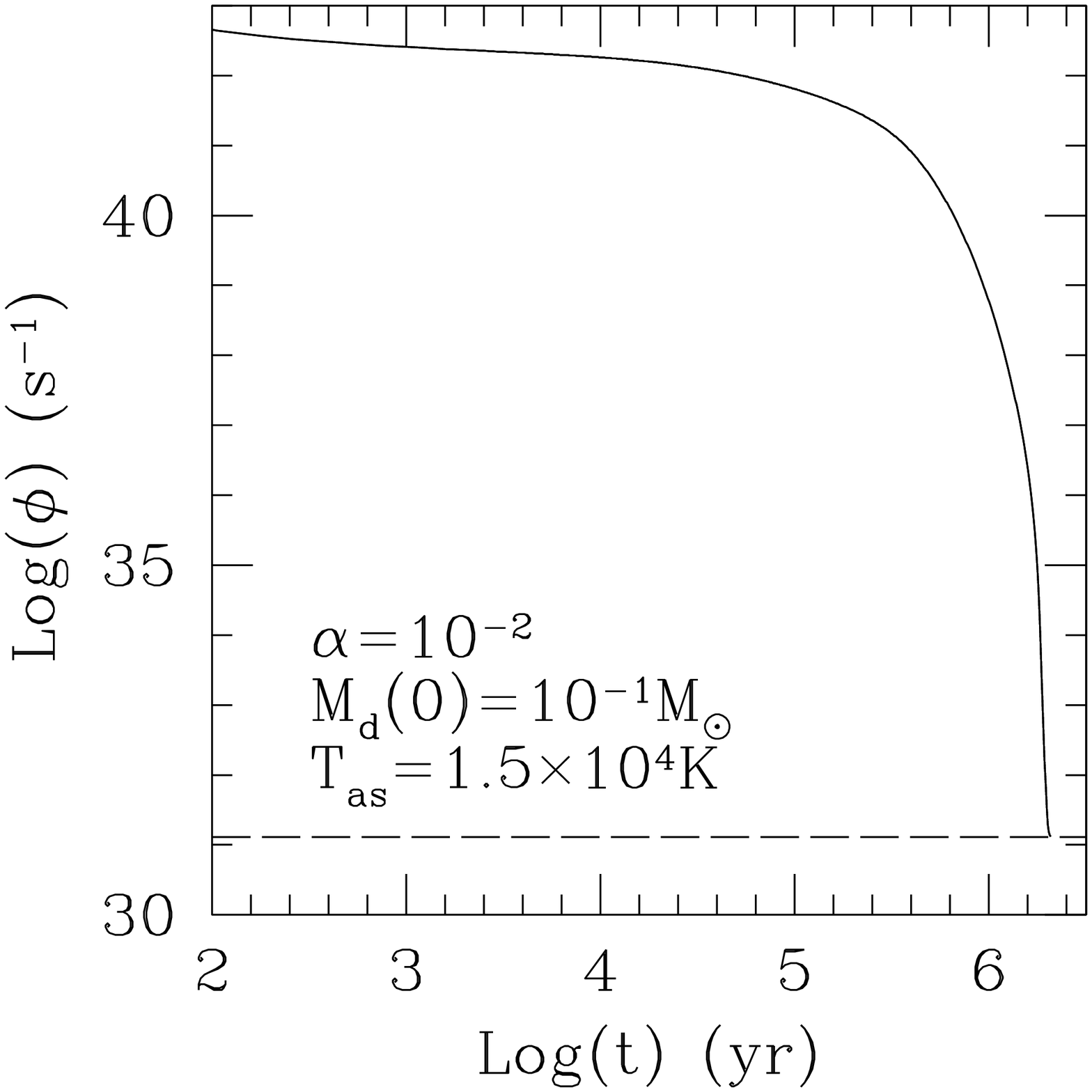}{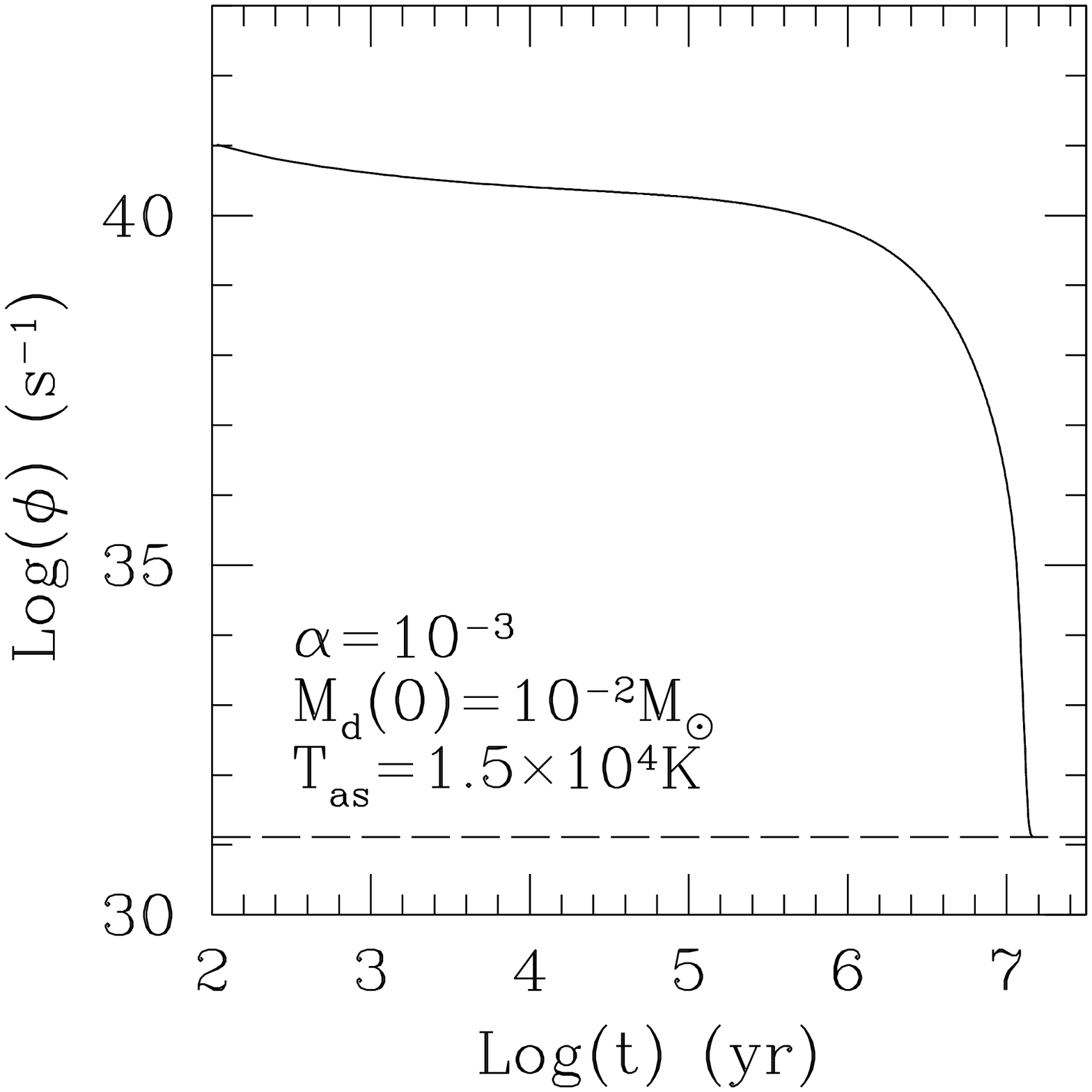}

\caption[]{Ionizing flux from the central source for disk removal
by viscous accretion, photoevaporation from the central source, and
FUV photoevaporation by external stars as a function of disk lifetime.
The ionizing flux decreases at late stages of the disk evolution with
the accretion rate. The long-dashed line indicates the constant ionizing
flux from the quiescent stellar photosphere (\( \phi _{\star
}=1.29\times 10^{31}\textrm{s}^{-1} \)). 

\label{centralFUV flux} }

\end{figure}

\begin{figure}

\plottwo{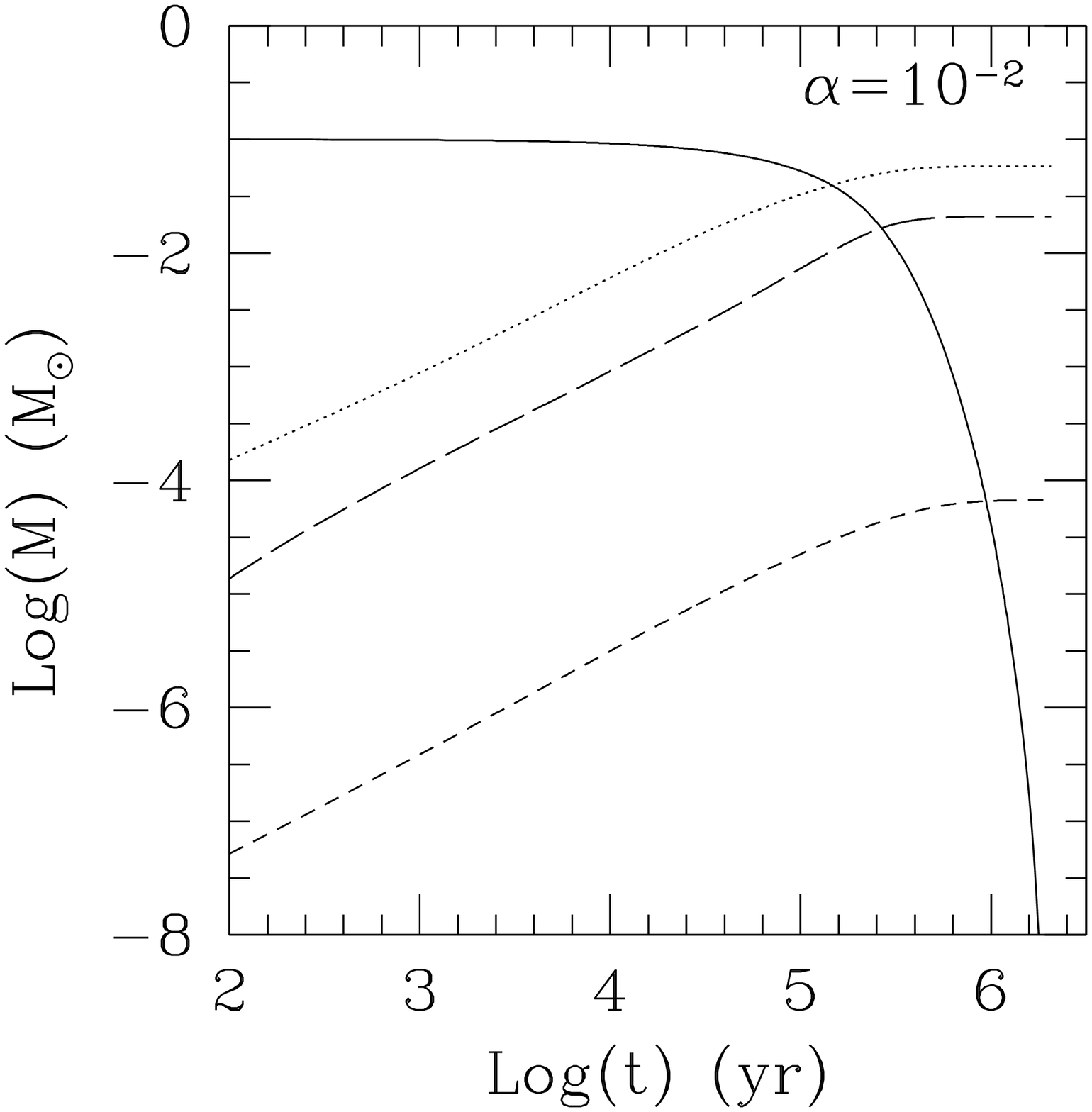}{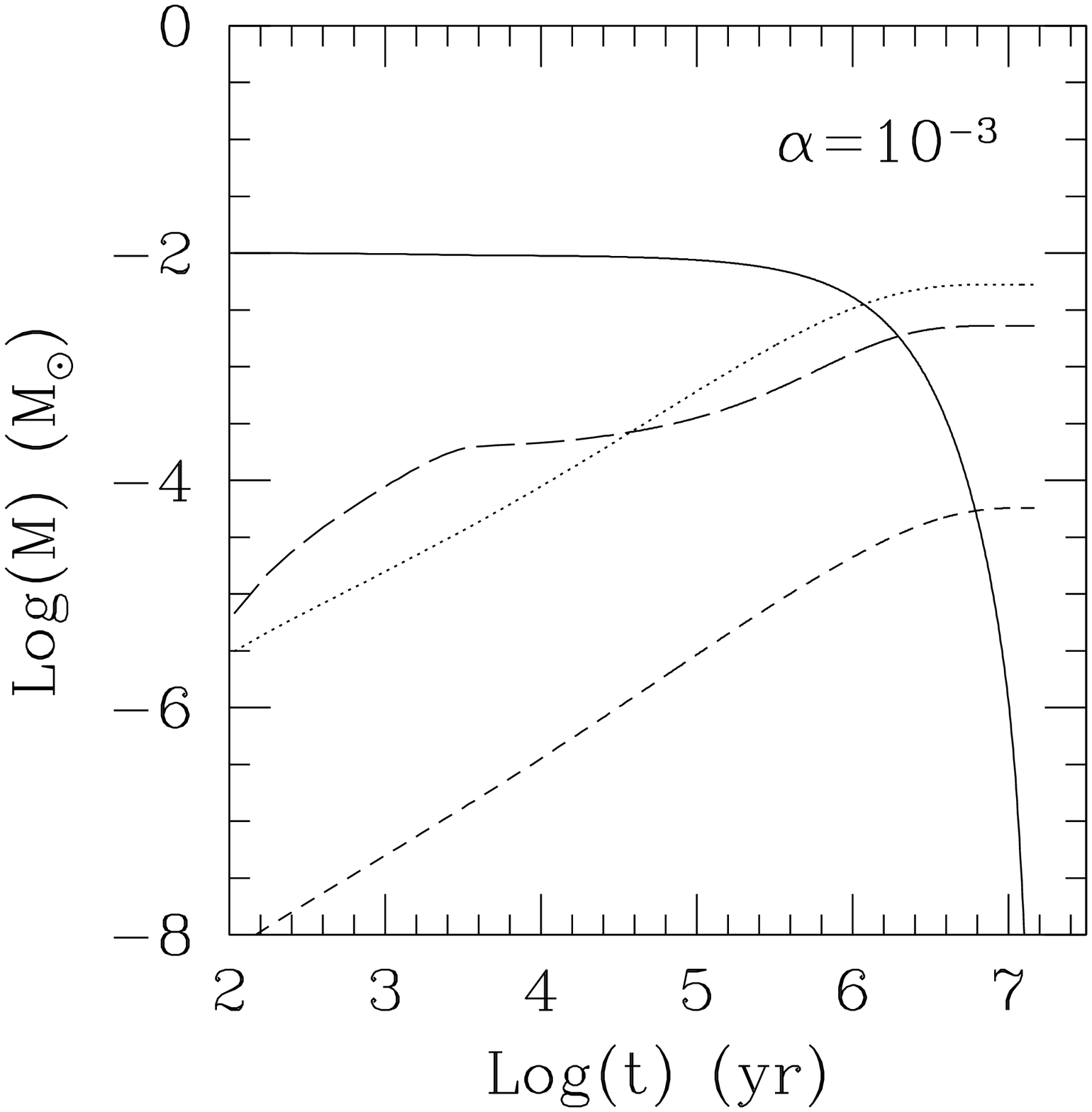}

\caption[]{Disk mass (solid line), disk mass accreted toward the central
star (dotted line), disk mass removed by photoevaporation from the
central source (short-dashed line), and disk mass removed by FUV
photoevaporation
by external stars (long-dashed line) vs. disk lifetime for the two
representative models. 

\label{centralFUV mass} } 

\end{figure}

We show the results for the two representative models with the same
parameters as before. Figure \ref{centralFUV sigma} shows snapshots
of the surface density evolution. The outer disk (\( R>R_{gI} \))
is removed by FUV photoevaporation from the external stars. As a result,
the disk edge is reduced to the FUV gravitational radius. The disk
size remains roughly constant at this radius until the disk is
completely
removed. At the final stages of the disk evolution (\( t\sim t_{gap}
\)),
when the surface density is very low (\( \Sigma \sim 10^{-5}\textrm{g
cm}^{-2} \)),
a gap structure is created by photoevaporation from the central source.
The gap formation is possible because photoevaporation from the central
source and photoevaporation by the external stars are dominant at
different locations (\( R_{gII} \) and \( R_{gI} \)). Both annuli
are removed as viscous accretion, viscous spreading, and
photoevaporation
work in resonance (see figures \ref{centralFUV mass} and \ref{centralFUV
mdot}).
The outer annulus is removed by the combination of viscous spreading,
photoevaporation from the central source, and photoevaporation by
the external stars. On the other hand, the inner annulus is removed
by the combination of viscous accretion toward the central star, viscous
spreading, and photoevaporation from the central source. The efficient
inner disk removal is self-consistent since we have calculated the
ionizing flux from the accretion luminosity and the quiescent stellar 
photosphere.
As the inner disk is removed, the contribution from the accretion
luminosity becomes negligible; as a result, the ionizing flux reaches
the value corresponding to the quiescent stellar photosphere, \( \phi
_{\star }=1.29\times 10^{31}\textrm{s}^{-1} \)
(figure \ref{centralFUV flux}). Again, the active chromosphere of these
low-mass stars has not been included; however, in this situation
the lack of chromospheric ionizing flux is much less important as
the timescale for disk evaporation has been set by the disk truncation
due to the external source.

\section{Discussion and summary}

We have studied the possibility of disk removal by the combination
of viscous diffusion and photoevaporation, assuming that the ultraviolet
photons responsible for evaporation arises either from the quiescent
stellar 
photosphere, the accretion shock, or external O stars. It is not
possible
to remove the entire disk when the only disk removal mechanism is
viscous accretion: the disk spreads indefinitely in order to conserve
total angular momentum while material is accreted onto the central
star. Mass loss due to photoevaporation removes material along with
its specific angular momentum; thus, it is possible to accrete material
toward the central star and reduce the amount of disk spreading. The
combination of the two mechanisms can result in finite disk lifetimes.

The distinctive features of photoevaporation by the ionizing flux from
the central source are the formation of a gap around the EUV
gravitational
radius at late stages of the disk evolution and the lack of
a finite time for the complete dispersal of the disk. The gap forms for
the
observed range of accretion shock temperatures (\( 1-3\times 10^{4} \)K)
and the disk becomes divided into an inner and an outer annulus. The
inner annulus continues to be removed by the combination of viscous
accretion,
viscous spreading of material beyond the EUV gravitational radius,
and photoevaporation at this radius. The outer
annulus is removed as viscous spreading of material toward the
gravitational
radius and photoevaporation work in resonance. \citet{uvswitch}
considered
models with photoevaporation by a constant central ionizing flux
combined
with viscous evolution, and showed that the timescales to remove the
inner 
and the outer annuli are not the same. We conclude that this result is
due 
to their assumption of a constant ionizing flux.
In contrast, we calculate the ionizing flux from the accretion
luminosity
self-consistently.  We find that both the inner and outer disks survive
to much longer times, and that the inner disk is {\it not} removed
first.
It is not possible to quickly remove the inner 
annuli and maintain a high ionizing flux at the same time because the 
accretion rate decreases.  The formal disk lifetime is found to be in
the 
range \( 10^{12}-10^{13} \)yr for \( 10^{-3}<\alpha <10^{-2} \) and 
\( 10^{-2}<M_{d}(0)<10^{-1} \), much longer than the star lifetime. 
Introducing a stellar chromosphere will produce models intermediate
between those presented here and those of \citet{uvswitch}. 

Stellar disk material can also be evaporated by external stars. This
is a likely situation for the disks surrounded by ionization fronts
in the Trapezium Cluster \citep{j98, sh99, orion observations}. The EUV
photons
dominate the flux for the proplyds orbiting at a distance \( d\la 0.03
\)pc,
and the FUV photons dominate for the proplyds at more typical (\( d\gg
0.03 \)pc)
distances from the massive external stars.

We consider photoevaporation due to both the central source and external
stars. There is no gap formation for disks nearby external hot stars
since their ionizing flux removes all the disk material outside the
EUV gravitational radius. The disk is quickly removed by viscous
accretion,
viscous spreading, and photoevaporation outside the gravitational
radius. The disk lifetime is in the range \( 10^{5}-10^{6} \)yr for
the same parameters as before, where the shorter lifetimes correspond
to shorter viscous evolution timescales of high viscosities. The disk
lifetimes are shorter by roughly two orders of magnitude when compared
to the models with only photoevaporation from the central source.
The short disk lifetimes are due to the strong influence of the external
radiation field at the outer disk, where most of the disk mass initially
resides.

The disk evolution is very different at larger, more typical distances
(\( d\gg 0.03 \)pc) from the external stars due to the different
locations where photoevaporation is efficient. The disk material in
the vicinity of the EUV gravitational radius is removed by
photoevaporation
from the central source. Similarly, the disk material in the
neighborhood
of the FUV gravitational radius is removed by external FUV
photoevaporation.
The effect of external photoevaporation is initially very dominant;
quickly reducing disk size to the FUV gravitational radius. After
this disk truncation, the disk mass is removed at two locations. At
the inner edge, disk material is viscously accreted onto the central.
On the other hand, viscous diffusion spreads the disk material beyond
the FUV gravitational radius, and FUV photoevaporation removes the
disk material at this location. The combination of this mechanisms
decreases the disk surface density. At the last stages of the disk
evolution, when the surface density is very low (\( \Sigma \sim
10^{-4}\textrm{g cm}^{-2} \)),
the ionizing from the central source opens a gap in the disk. In this
case, it is possible to remove the inner annulus faster than the outer
one self-consistently. The disk lifetime is in the range \(
10^{6}-10^{7} \)yr
for the same parameters as before, where the shorter disk lifetimes
correspond to the shorter viscous evolution timescales of high
viscosities. 

Observations suggest that the timescale for all the stars in young
clusters to lose their disks is \( \sim 6\times 10^{6} \)yr
\citep{lifetime=6Myr},
in agreement with the disk lifetimes for the proplyds nearby external
hot stars (\( 10^{5}-10^{6}\textrm{yr} \)). The disk removal timescale
is an important constraint on the timescale allowed for planet
formation.
It is very attractive to study the possibility of planet formation
in star forming regions since this process follows star and disk
formation.
If the influence of massive external stars in star forming regions
is not demolishing for planet formation, star forming regions are
natural birthplaces of extrasolar planets. Numerical simulations suggest
a timescale for terrestrial planet formation in the range \(
10^{7}-10^{8} \)yr
\citep{planet formation terrestrial planets 8, planet formation 20 Myr
for terrestrial planets},
and a typical timescale for giant planet formation that range from
a few million years to more than \( 10^{7} \)yr \citep{giant planet
formation, giant planet formation > 8}.
It is thus not possible to form planets around stars in the neighborhood
of massive O stars (i.e. EUV external photoevaporation) because the
disk lifetime is too short (\( 10^{5}-10^{6}\textrm{yr} \)). On the
other hand, at typical distances from the external stars (\( d\gg 0.03
\)pc),
the disk lifetimes are long enough (\( 10^{6}-10^{7}\textrm{yr} \))
to allow for the formation of terrestrial and giant planets. There
are no constraints on planet formation in the absence of EUV or FUV
fluxes from external stars.

\acknowledgements{}

We would like to thank Norman Murray and Gibor Basri for helpful
comments and discussions. 
The research of D.J. and I.M. was supported through a grant from the
Natural Sciences and Engineering Council of Canada. The research of
I.M. was supported by the Connaught Scholarship of the University
of Toronto.

\clearpage

\begin{figure}

\plottwo{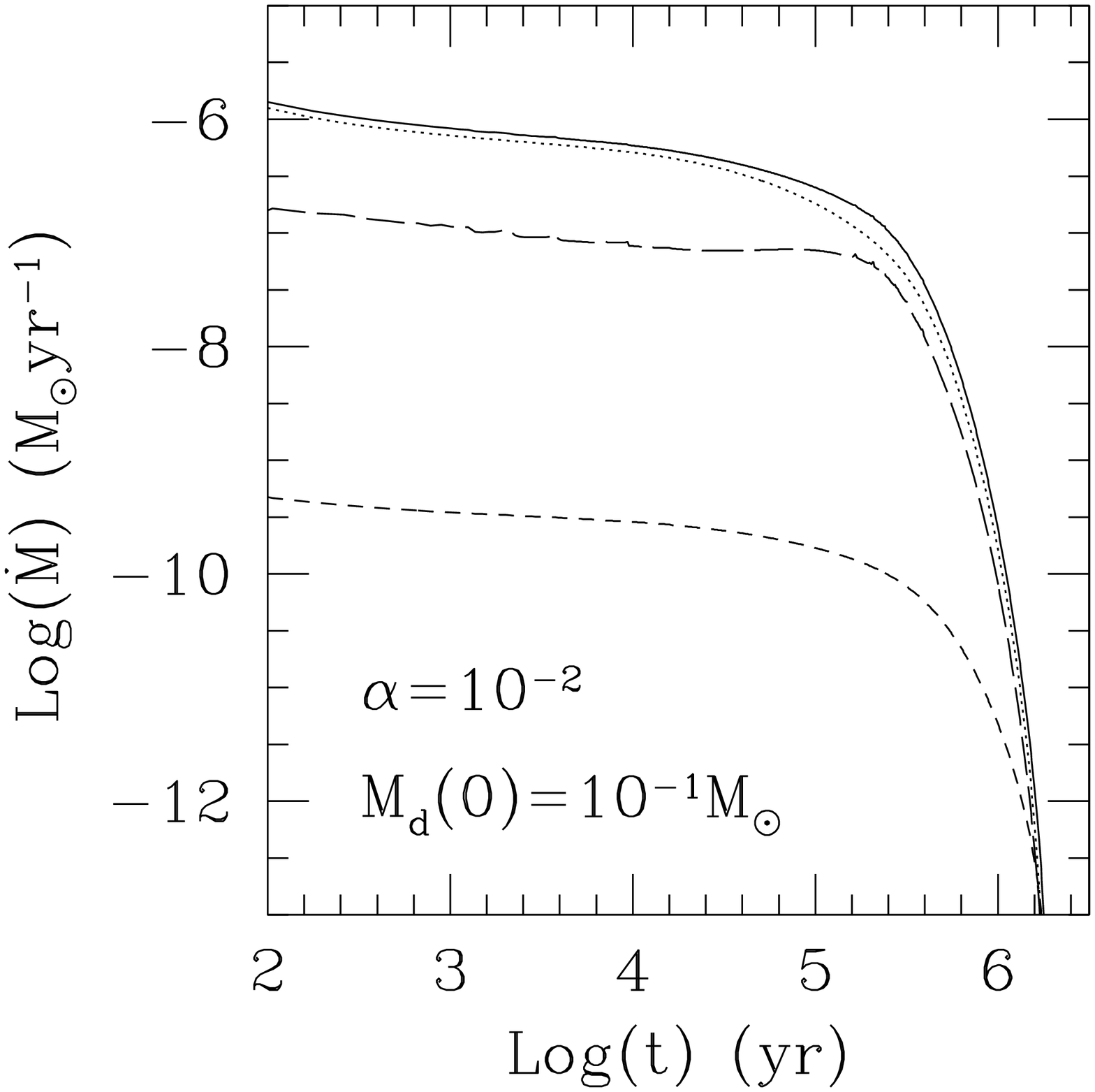}{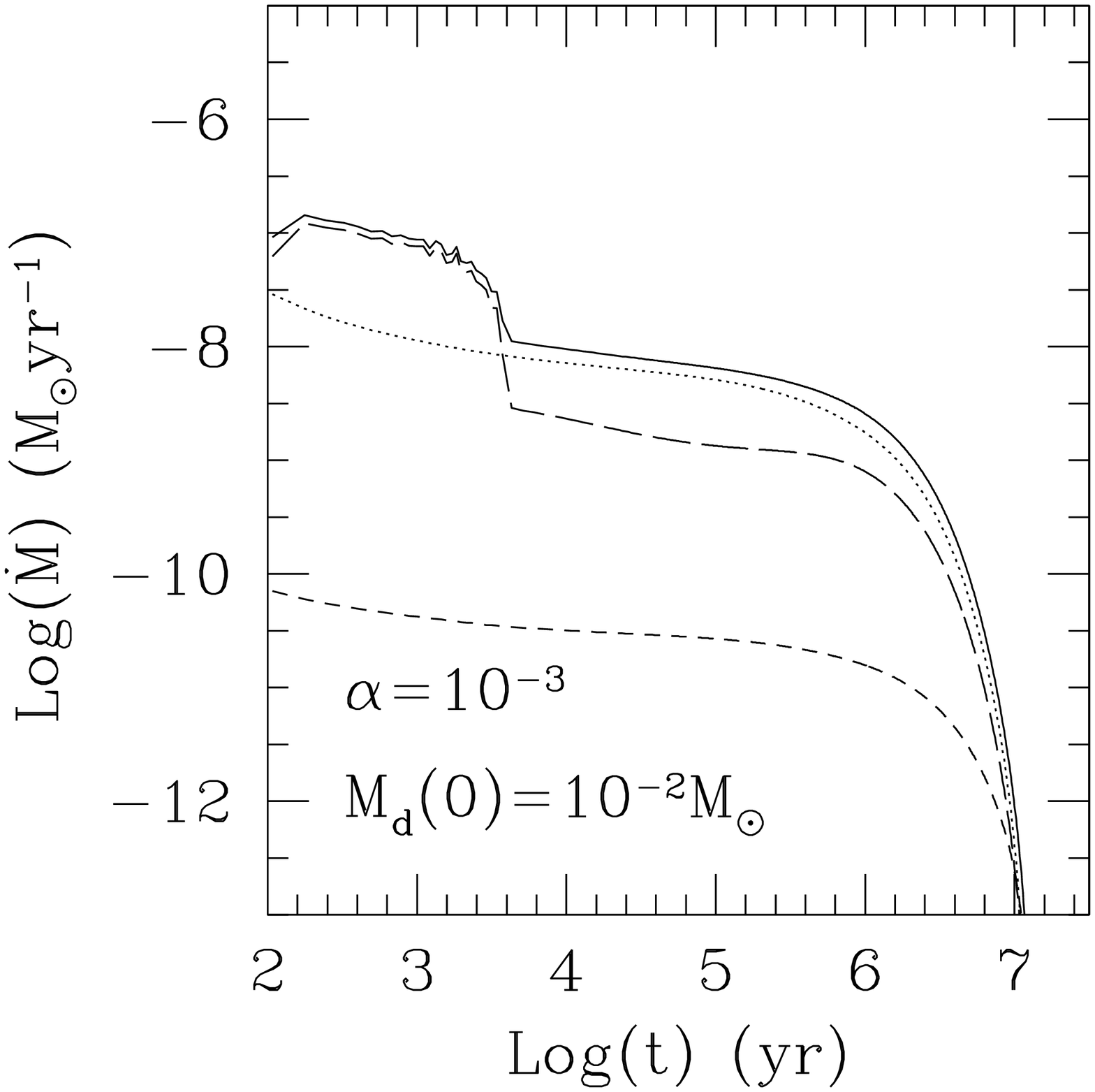}

\caption[]{Total mass removal rate (solid line), mass removal rate
due to accretion (dotted line), mass removal rate due to
photoevaporation
by the central source (short-dashed line), and mass removal rate by
the external stars (long-dashed line) as a function of disk lifetime for

the two representative
models. The disk is removed by viscous diffusion, photoevaporation
from the central source, and FUV photoevaporation by external stars.

\label{centralFUV mdot} }

\end{figure}
\end{document}